\documentclass{cecs}
\usepackage[latin9]{inputenc}
\usepackage[colorlinks, citecolor =blue,linkcolor=blue]{hyperref}

\def\mH{{\mathcal H}}
\def\pa{\partial}

\def\L{{\cal L}}
\def\O{{\cal O}}

\usepackage{float}

\newcommand{\be}[1]{\begin{equation}\label{#1}}
\newcommand{\cit}[1]{[\citenum{#1}]}
\newcommand{\ee}{\end{equation}}

\usepackage{mathtools}

\newcommand{\mbf}[1]{\boldsymbol{#1}}

\begin{document}







\toctitle{Regge-Teitelboim analysis of the symmetries of electromagnetic and
gravitational fields on asymptotically null spacelike surfaces}
\tocauthor{C. Bunster, A. Gomberoff, A. Perez}


\chapter[]{Regge-Teitelboim analysis of the symmetries of electromagnetic and gravitational fields on
 asymptotically null spacelike surfaces\raisebox{.2\baselineskip}{\normalsize\footnotemark}}\footnotetext{To appear in the forthcoming volume ``Tullio Regge: an eclectic genius, from quantum gravity to computer play,'' Eds. L. Castellani, A. Ceresola, R. D'Auria and P. Fr\'e (World Scientific).}


\vskip 0.5cm
\begin{flushright}

\emph{To Tullio, for old times' sake}

\end{flushright}

\author{Claudio Bunster$^\dagger$, Andr\'es Gomberoff $^{*,\dagger}$ and Alfredo P\'erez$^\dagger$}
\address{${}^\dagger$Centro de Estudios Cient\'{\i}ficos (CECs), Avenida Arturo Prat 514, Valdivia, Chile \\ ${}^*$Facultad de Ingenier\'ia y Ciencias, Universidad Adolfo Ib\'a\~nez, Avda.~Diagonal las Torres 2640, Pe\~nalol\'en, Santiago, Chile}

\begin{abstract}
We present a new application of the Regge-Teitelboim method for treating
symmetries which are defined asymptotically. It may be regarded as
complementary to the one in their original 1974 paper. The formulation
is based on replacing an asymptotic plane by the two--sheeted ``hourglass" shaped
 surface obtained by joining smoothly an incoming hyperboloid with an outgoing one.
  The hyperboloids have a fixed radius, and as one moves the center of the hourglass
   along the time axis one covers the whole of spacetime. The motivation is to study radiation, and
    the hourglass is well suited to the
task because it is asymptotically null, and thus is able to register
the details of the process. A simple parity condition for the fields on the hyperboloid is given. It specifies that as much radiation as is coming in as it is going out. With it, a Hamiltonian
formulation of the symmetry of Bondi, van der Burg, Metzner and Sachs is developed fir both electromagnetism and gravitation. It is indispensable for the construction to have electric--magnetic duality asymptotically. For gravitation, a formulation for the linearized theory on the hourglass has not been explicitly constructed; but enough rudiments of it are given so that the main results can be established. A definition for angular momentum wish is conserved (for which the ``magnetic sector" is essential) is given. It incorporates an interrelationship between spin and charge. For the gravitational field, Taub-NUT space appears as the analog of a magnetic pole.

\end{abstract}
\noindent

\section{Introduction}


The 1974 paper by Regge and Teitelboim [\citenum{Regge:1974zd}] contained
two main results: (i) A completion of Dirac's analysis [\citenum{Dirac:1958sc}]
of the role of constraints in field dynamics that was necessary in
order to account for the different character of the gauge transformations
which do not change the physical state (``proper gauge transformations''),
from those which do (``improper gauge transformations''). It was
found that, in the latter case, Dirac's ``weakly vanishing'' generators
have to be improved by the addition of a surface integral and do not
vanish weakly. The surface integral gives the value of the charge
associated to the improper transformations. (ii) An application of
(i) to obtain a Poincar\'e invariant formulation of the theory of gravitation
on spacelike surfaces which are asymptotically planes.

       We develop below a new application of the method which is based on replacing an asymptotic
     plane by the two--sheeted ``hourglass" shaped
 surface obtained by joining smoothly an incoming hyperboloid with an outgoing one.
  The hyperboloids have a fixed radius, and as one moves the center of the hourglass
   along the time axis one covers the whole of spacetime. The motivation is to study radiation, and
    the hourglass is well suited to the
task because it is asymptotically null, and thus is able to register
the details of the process. In contradistinction, if one goes far enough on an asymptotic plane,
 the radiation emitted by a confined source has not enough time to reach there.
  
 A simple parity boundary condition, with a direct physical meaning,
 for the fields defined on the hourglass is given. With it, a Hamiltonian
formulation of the symmetry of Bondi, van der Burg, Metzner and Sachs
 \cit{Bondi:1962px,Sachs:1962zza,Sachs:1962wk} is developed. The formulation
yields a Hamiltonian
    definition of the Bondi ``news''. For electromagnetism, an interrelationship between spin and
     the BMS charge is brought out, and manifest electric-magnetic duality is achieved by 
     introducing a second vector potential. For the gravitational field, Taub-NUT space appears as a gravitational
      magnetic pole and  the magnetic
    BMS charges are exhibited. However, and expression for their conjugates, the 
   magnetic supertranslations, is lacking. 
   
   The ``hourglass" construction, the parity conditions on it and the intimate connection between BMS invariance and electric--magnetic duality were reported in \cit{short}

If one gives initial data on one hyperbolic hourglass, the Hamiltonian equations of motion determine
 the canonical variables on any other hyperbolic hourglass. However, the initial data should only be 
 specified on
  one half of the 
  hourglass. This is because a point, which is not at infinity, lying, say, on the outgoing half of one hourglass
   at a given time also lies on the incoming half of another hourglass at a later time. This double counting
    does not happen at infinity, and hence one needs to specify the radiation 
    on the other half. That is, if one is giving the data on the outgoing half one must specify in addition
     the incoming  Bondi news. But the parity condition states that they are equal to the outgoing 
    ones, and therefore it is sufficient to specify just the data on the outgoing half of the hourglass
    (or, viceversa, on the incoming one). In this sense the
    hyperbolic hourglass  is a Cauchy surface.

If compared with an hourglass formed by incoming and outgoing light cones,
the one formed by hyperboloids has the advantage of being spacelike and therefore
permitting direct step by step use of the Dirac's procedure with the
Regge-Teitelboim complement, which has been battle-tested, and in
which all the structures that appear (action, Hamiltonian, Poisson
and Dirac brackets, surface deformations, most general permissible
motion) are well, and tightly, defined from the start.

Furthermore, the slicing by hyperbolic hourglasses of fixed radius and varying
center has the essential property of covering the whole
of spacetime, in contradistinction with the foliations by hyperboloids
of fixed center and varying radius, used previously by many authors,
which only cover part of it.

The structure of the paper is the following. In order to make the
treatment self-contained and set the notation and terminology, section \ref{secII} begins by reviewing the general procedure. Section \ref{foliation}
discusses the foliation by hyperbolic hourglasses of the same radius and different
center. Next, section \ref{sec:EMMink} contains an analysis
of the asymptotic properties of the free electromagnetic field on
the hyperbolic hourglass foliation. The discussion is given in detail, because
practically all the results derived for electromagnetism can be translated
literally to the gravitational case, whose treatment becomes then considerably lighter. Section \ref{sec:Gravfield} is then devoted to the gravitational case.

Three appendices
are included: appendix \ref{app1} gives explicit expressions for
the Poincar\'e generators on the hyperbolic hourglass foliation. Appendix \ref{app2} 
gives the details of the preservation in time of the parity boundary conditions. 
Appendix \ref{sec:app3} provides a ``dictionary'' for translating
in the gravitational case the variables which appear in the present
Hamiltonian treatment with those employed in the original BMS light
cone analysis.


\section{Hamiltonian field dynamics, surface deformations, gauge transformations,
surface integrals, and conservation laws.}

\label{secII}

\subsection{Quick review of formulation.}

\label{qrf}

In the formulation of field dynamics in which the state is defined
on a general spacelike surface developed by Dirac \cit{Dirac:1958sc},
and completed by Regge and Teitelboim \cit{Regge:1974zd} to incorporate
symmetries which are defined asymptotically, the generator\textendash through
Poisson brackets\textendash of the ``most general permissible motion''
has the form 
\begin{equation}
H[\xi;\lambda]=H_{0}[\xi;\lambda]+Q[\xi;\lambda]\,,\label{impr}
\end{equation}
where $H_{0}[\xi;\lambda]$ is an integral over the spacelike surface
on which the state is defined, of the form 
\begin{equation}
H_{0}[\xi;\lambda]=\int d^{3}x\left(\xi^{\mu}\mH_{\mu}+\lambda^{a}G_{a}\right)\,,\label{Ham}
\end{equation}
and $Q[\xi,\lambda]$ is a surface integral over the asymptotic boundary
of that spacelike surface.

The surface integral $Q[\xi;\lambda]$ is included to make well defined
the functional derivatives of $H[\xi;\lambda]$, so that one has,
\begin{equation}
\delta H[\xi;\lambda]=\int d^{3}x\left(\frac{\delta H}{\delta\phi}\delta\phi+\frac{\delta H}{\delta{\pi}}\delta\pi\right)\,,\label{dG}
\end{equation}
\textit{without any surface terms}. This means that the contribution
of the asymptotic part of the field is already included in \eqref{dG}.
Here we have abbreviated as $(\phi,\pi)$ all the canonical field
variables of the theory.

In \eqref{Ham} the $\mH_{\mu}$ are the generators of deformations
of the spacelike surface in which the state is defined, while the
$G_{a}$ are the generators of internal gauge transformations. They
are both constrained to vanish, that is, they are weakly equal to
zero: 
\begin{equation}
\mH_{\mu}\approx0,\ \ G_{a}\approx0\,.\label{constraints}
\end{equation}

\subsubsection{Proper and improper gauge transformations}

If the parameters $(\xi,\lambda)$ are such that the surface integral
$Q[\xi;\lambda]$ vanishes, the motion generated by $H_{0}[\xi;\lambda]$
is called a ``proper'' gauge transformation \cit{Benguria:1976in},
and it it is not a symmetry, but rather an expression of the fact
that the system is described by variables which are redundant. This
normally happens when they vanish at infinity, but there are important
cases, in both electromagnetism and gravity \cit{Regge:1974zd} where
the surface integrals vanish even though the parameters do not vanish
at infinity but obey parity conditions there. In that case the transformation
is still proper. For an internal symmetry one feels on safe grounds
stating that a proper gauge transformation does not change the physical
state on a given spacelike surface. For the case of a surface deformation
this point of view may be kept for purely tangential deformations
(changes of spatial coordinates), but if the deformation has a normal
component the hypersurface is geometrically deformed, and therefore,
if one can ``perform local observations inside\char`\"{}, one would
expect the physical state to change. On the other hand if one only
performs observations at infinity, then one may safely take the point
of view that proper normal deformations are also gauge transformations,
and do not change the physical state either. If $(\xi,\lambda)$ are
such that the surface integral $Q[\xi;\lambda]$ does not vanish the
motion is called an ``improper\char`\"{} gauge transformation, and
one expects it to change the physical state. As a consequence of \eqref{dG},
the functional form of the transformation generated by \eqref{impr}
is the same if the transformation is proper or improper, the difference
is only introduced by the asymptotic behavior of the transformation
parameters.

The difference between proper and improper transformations manifests
itself at the level of the action principle in which \eqref{impr}
is the Hamiltonian, in the fact that the constraints \eqref{constraints}
are obtained from it by extremizing the action with respect to $\xi$
and $\lambda$ \textit{keeping fixed the part that contributes to
the surface integral $Q[\xi,\lambda]$}. Thus, \eqref{constraints}
states that the generator of proper gauge transformations vanishes
weakly. On the other hand, for improper gauge transformations this
does not happen: 
\[
H[\xi;\lambda]\approx Q[\xi;\lambda]\neq0\,.
\]
If one is interested in the action of an asymptotic motion, one gives
the asymptotic part of the transformation and continues it inside
in an arbitrary manner. The way in which one chooses to continue inside
is irrelevant because any two continuations differ by a proper gauge
transformation. It is however necessary to continue, because it is
only the sum of the volume part of the generator and the surface term
which has well defined functional derivatives and is therefore capable
of acting through a Poisson bracket. Neither the volume part alone,
nor the surface integral alone has that capability. Alternatively,
one may choose a particular continuation inside by fixing the gauge.
If this is done, the combined set of the original gauge constraints
and the gauge conditions become second class, and one can pass from
the original Poisson bracket to the associated Dirac bracket, in terms
of which the second class constrains vanish strongly and have zero
bracket with everything. Then the surface term stands alone and is
capable to act as itself as a generator through the Dirac bracket.

\subsubsection{Commutation of deformations}

The constraint\textendash generators $\mH_{\mu}$, $G_{a}$ are first
class. If we denote collectively by $\Lambda_{1}=(\xi_{1},\lambda_{1})$,
$\Lambda_{2}=(\xi_{2},\lambda_{2})$, the parameters of any two motions
one has, for proper gauge transformations, 
\begin{equation}
\Big[H[\Lambda_{1}],H[\Lambda_{2}]\Big]=H[\Lambda_{12}]\,,\label{pbH}
\end{equation}
where the commutator $\Lambda_{12}=-\Lambda_{21}$ of the two original
infinitesimal transformations is a bilinear expression of $\Lambda_{1}$,
$\Lambda_{2}$ and their derivatives (in practice, first derivatives,
with coefficients which in general depend on the fields). For improper
transformations it may happen that Eq. \eqref{pbH} is relaxed by
the appearance of a central extension on its right hand side, that
is, by the addition to $H[\Lambda_{12}]$ of a term that has zero
Poisson brackets with all the dynamical variables. This is not allowed
for proper transformations because it would spoil the first class
character of the constraints; i.e. the vanishing of the constraints
would not be preserved by the transformation. This obstruction to
the presence of a central extension does not happen for improper transformations
because the charge $Q$ is not constrained to be zero.

Given any theory one may calculate $\Lambda_{12}$ by working out
directly the Poisson bracket on the left side of \eqref{pbH}. However,
if one has geometrical insight on the nature of the motions at hand
one may write down the result without doing that calculation. For
example, for a Yang-Mills theory, with structure constants $C_{\ bc}^{a}$,
one has 
\[
\lambda_{12}^{c}=C_{\ ab}^{c}\lambda_{1}^{a}\lambda_{2}^{b}\ ,
\]
and for two surface deformations within an arbitrary Riemannian spacetime
with Lorentzian signature one has \cit{Teitelboim:1973yj,Teitelboim:1972vw},
\begin{eqnarray}
\xi_{12}^{\perp} & = & \xi_{1}^{i}\xi_{2,i}^{\perp}-\xi_{2}^{i}\xi_{1,i}^{\perp}\,,\label{def1}\\
\xi_{12}^{i} & = & g^{ij}\left(\xi_{1}^{\perp}\xi_{2,j}-\xi_{2}^{\perp}\xi_{1,j}\right)+\xi_{1}^{j}\xi_{2,j}^{i}-\xi_{2}^{j}\xi_{1,j}^{i}\,,\label{def2}
\end{eqnarray}
where $g_{ij}$ is the metric of the spacelike surface.

\subsubsection{No royal road}

Lastly, we elaborate on the sentence ``...The surface integral $Q[\xi;\lambda]$
is included in \eqref{impr} to make well defined the functional derivatives
of $H[\xi;\lambda]$...\char`\"{} written above. To achieve this it
is necessary to find an appropriate set of boundary conditions. There
is no foolproof, inductive method for that. One rather works by trial
and error and there is no guarantee of success. The procedure in practice
is as follows: (i) A tentative set of asymptotic conditions is obtained
by applying the asymptotic transformations that one wants to have
present, to a simple field configuration that one also wants to have
present. For example, in the case of gravitation, one would boost
a Schwarzschild field; or, in electromagnetism, a Coulomb field. (ii)
One extracts properties of the result obtained that can be formulated
independently of the specific original configuration, and uses them
as a starting ansatz. For example, one may retain a decay rate in
inverse powers of the radial distance and a parity condition for the
coefficients.(iii) One finds the most general parameters $(\xi,\lambda)$
which preserve the ansatz. It may happen then that that set of parameters
does not contain all the symmetries that one was interested in (for
example, the complete Poincar\'e group). Then one relaxes
the ansatz to make room. If success is achieved \textendash{} meaning,
in the example just given, that one has the Poincar\'e group or more \textendash{}
one checks whether the surface integral that appears in the variation
of the $H_{0}$ is the variation of a finite surface integral, or,
as one says colloquially, ``if the $\delta$ can be taken out\char`\"{}.
If this happens one is done. If it does not, one modifies the boundary
conditions in the light of the nature of the failure. With luck and
dedication the process converges and one finally succeeds.

\section{Foliation of Minkowski space by hyperboloids
of the same radius and different centers}

\label{foliation}

\subsection{Incoming and outgoing hyperboloids}

In Minkowski space it is natural to define the state on a three-dimensional
spacelike hyperboloid, because that surface is mapped onto itself
by a Lorentz transformation. In this sense, hyperboloids are more
adequate to the special principle of relativity than the planes corresponding
to inertial frames, for which the boosts are interchanged with spatial translations. This possibility was considered by Dirac
in 1940 \cit{RevModPhys.21.392} and he called it the ``point form
of field dynamics'' with the term ``point'' referring to the center
of the hyperboloid. 

A spacelike hyperboloid with center at $x_{(0)}^{\mu}$ and radius
$\tau_{0}$ obeys the equation 
\begin{equation}
(x^{\mu}-x_{(0)}^{\mu})(x_{\mu}-x_{(0)\mu})=-\tau_{0}^{2}\,.\label{hyp}
\end{equation}
Actually, Eq. \eqref{hyp} describes two disjoint hyperboloids, one
with $x^{0}>x_{(0)}^{0}$ (``outgoing hyperboloid'') and another
with $x^{0}<x_{(0)}^{0}$ (``incoming hyperboloid''). Although Dirac did not discuss foliations of spacetime by
means of a family of hyperboloids, this has been done by many authors
but to our knowledge in all cases treated so far the foliation has
been defined by keeping $x_{(0)}^{\mu}$ fixed and letting $\tau_{0}$
vary as one passes from one hyperboloid to the next.

In other words, the foliations used previously have consisted of a
sequence of hyperboloids with fixed center and varying radius\footnote{See, for example,
\cit{Ashtekar:1978zz}, \cit{0264-9381-9-4-019},  \cit{Campiglia:2015qka},
and also  \cit{Troessaert:2017jcm}\textcolor{red}{{} }and references
therein. In some of these discussions timelike hyperboloids are employed
(in which \nopagebreak case $-\tau_{0}^{2}$ in \eqref{hyp} is replaced by $\lambda_{0}^{2}$).}. These
foliations have the advantage that the four $x^{\mu}$ are treated
on the same footing so Lorentz invariance is manifest; but the price
payed is extremely high, because only a small part of Minkowski space
is covered, and moreover, a spurious explicit dependence on the varying
$\tau_{0}$, which is taken as the time $\tau$, is introduced. Here
we take the other natural option, we keep the radius $\tau_{0}$ fixed
and we allow the position of the center to vary. This is a direct
extension of what is done with null foliations, which may be regarded
as being the limit $\tau_{0}\rightarrow0$. The actual value of $\tau_{0}$
will turn out to be irrelevant, since all the quantities of physical
interest will incorporate naturally $\tau_{0}$ in their units.

\subsection{The hyperbolic hourglass}

The hyperbolic hourglass consists of an outgoing hyperboloid with center $x^\mu_{(0)}=(-\tau_0,0)$ joint
 to an incoming one  with center $x^\mu_{(0)}=(\tau_0,0)$.  At time $t$ it has spatial
  coordinates $(r,\vartheta,\varphi)$, whose range is:
\begin{eqnarray*}
	-\infty<&r&<+\infty, \\ 0\leq &\vartheta&\leq\pi, \\ 0\leq &\varphi&<2\pi.
\end{eqnarray*}
It is defined parametrically from Minkowskian coordinates $(x^0,\vec x)$ through
\begin{eqnarray}
x^{0} & = & t+ r\sqrt{1 + \frac{\tau_0^2}{ r^2}} + \tau_0  \ \ \ r\leq 0 \label{emb1}\\
x^{0} & = & t+ r\sqrt{1 + \frac{\tau_0^2}{ r^2}} - \tau_0    \ \ \ r\geq 0, \label{emb2}
\end{eqnarray}
and
\be{x}
\vec{x} = r \hat{r}\ , 
\  \ -\infty<r<+\infty,
\ee
where the unit vector $\hat r$ is given by
\be{unitr}
\hat r = (\cos\vartheta\cos\varphi,\cos\vartheta \sin\varphi, \sin\vartheta).
\ee
The radius $\tau_0$ is taken to be positive.  The embedding defined by the above equations is continuously differentiable. The tangent vectors are continuous at $r=0$ and the surface has a well defined global orientation. 

The hyperbolic hourglass may be regarded as a spacelike deformation of the full (pass and future) lightcone, with an orientation inherited from the propagation of a light front that comes in, goes through itself, and then comes out. Since this wave propagation process is physically smooth, fields defined on the global coordinate system just described should be smooth. (The parametric equations   \eqref{emb1}-\eqref{x} automatically incorporate the antipodal map \cit{Strominger:2013jfa} \cit{Strominger:2017zoo},  which amounts to rewriting them by using a positive $r$ for both sheets of the hyperboloid and inverting the orientation of the two-spehere at a given $r$.  That is, keeping   \eqref{emb1}-\eqref{x} for $r\geq 0$ and setting,
	$r' = -r$, $\hat r' = -\hat r$, for $r\leq 0$.)
	
	If one considers an incoming wave which is not spherically symmetric, then the spacetime point at which the wavefront goes through itself will be different for different $\hat r$'s. But in the present paper we are only interested in the analysis of the asymptotic region and therefore the details of what happens inside are irrelevant. The key aspects are the asymptotic hyperbolic shape and its orientation inherited from that of an incoming wave that goes through itself and becomes outgoing.

	Figure 1 shows the embedding in  Minkowski space of a single hyperbolic hourglass, figure 2 exhibits the slicing of Minkowski space by a one parameter family of hyperbolic hourglasses and figure 3 shows a sequence of Penrose diagrams with hyperbolic slicings of different radius $\tau_0$.

\begin{figure}[H]\label{hourglassfig}
\begin{center}
\includegraphics[width=10cm]{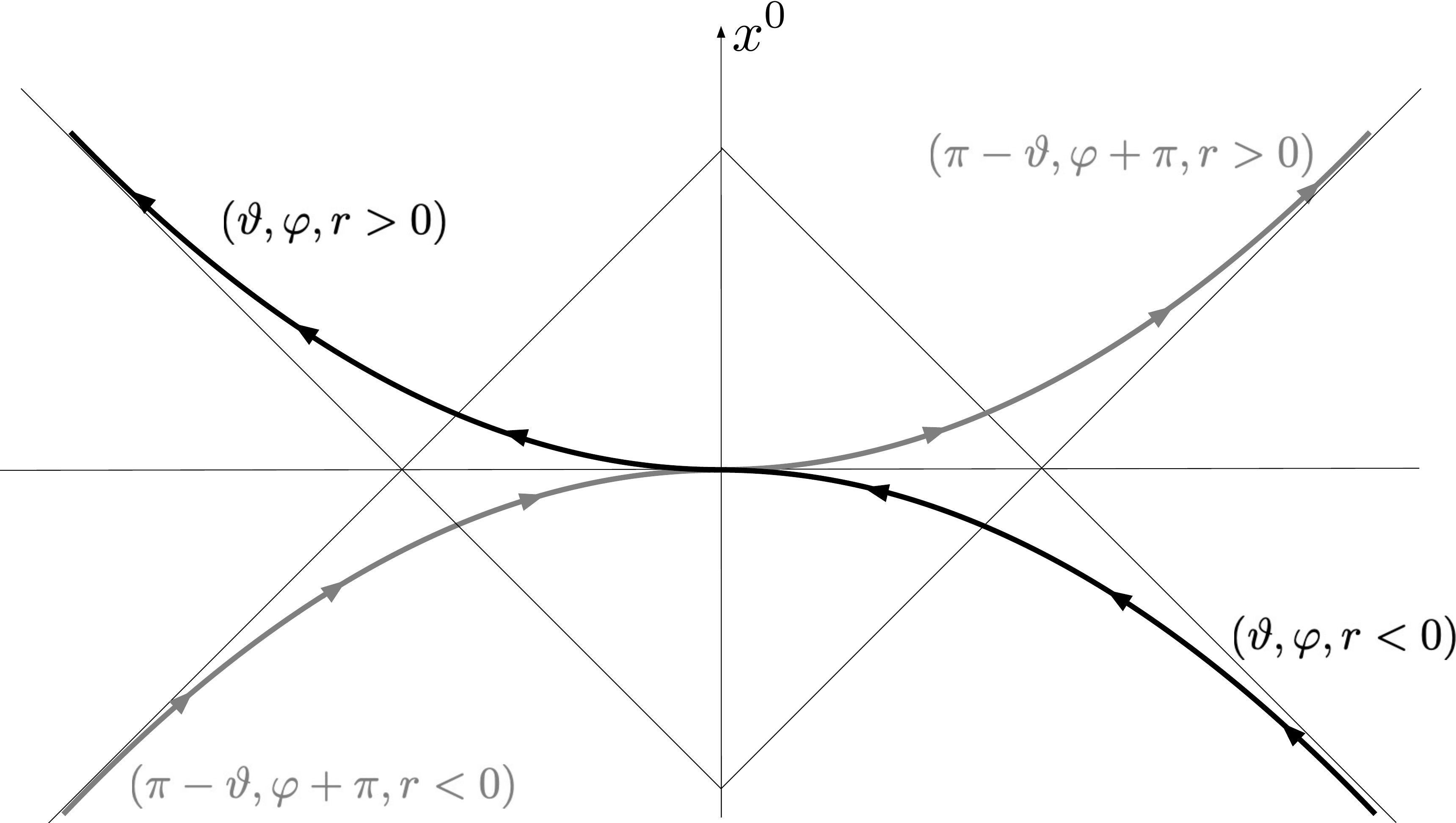}
\caption{{\it The hyperbolic hourglass}.
The figure shows a two dimensional cut of an incoming hyperboloid and an outgoing one which are joint smoothly at at $r=0$. The arrows show the direction of increasing $r$, which coincide asymptotically with the direction of propagation of a wave that comes in, goes through itself, and then comes out.
 If the incoming wave is not spherically symmetric, the spacetime point at which the wavefront goes through itself will be different for different $(\vartheta,\varphi)$. For the analysis of the asymptotic region the details of what happens inside are irrelevant. The key aspects are the asymptotic hyperbolic shape and its orientation inherited from that of an incoming wave that goes through itself and becomes outgoing.}
\end{center}
\end{figure}

\begin{figure}[H]\label{twopa}
\begin{center}
\includegraphics[width=9cm]{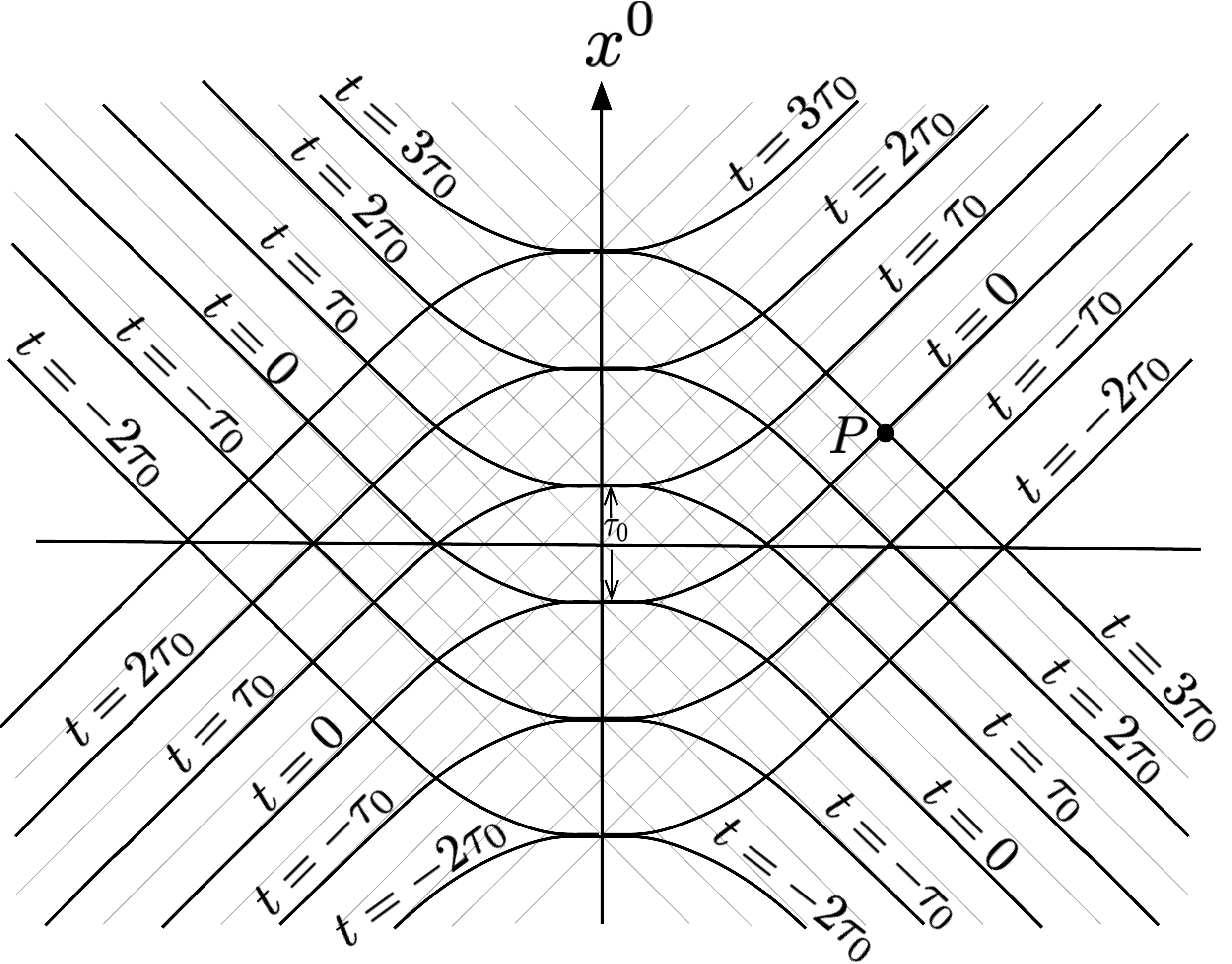}
\caption{{\it Slicing by hyperbolic hourglasses}. A given spacetime point is labeled by two set of
 coordinates. In the case of the point $P$ shown in the figure, these are $(t=0, r, \vartheta,\varphi)$ 
 and $(t=3\tau_0, -r, \pi-\vartheta,\varphi+\pi)$.}
\end{center}
\end{figure}
\begin{figure}[H]
\begin{center}
\includegraphics[width=13.5cm]{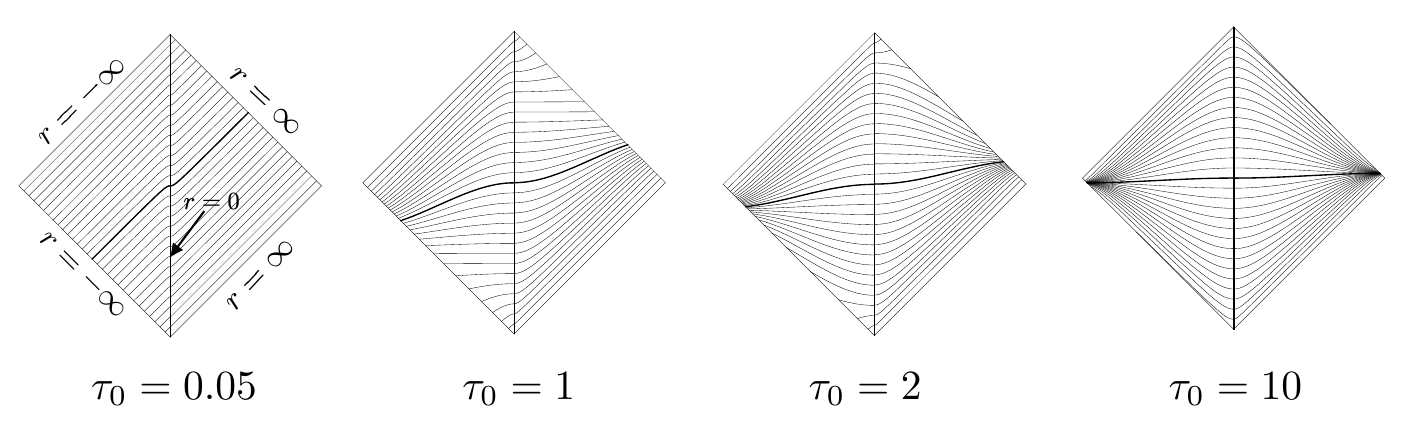}\label{mat}
\caption{{\it Limits $\tau_{0}\rightarrow0$ and $\tau_{0}\rightarrow \infty$ for Minkowski space}.
The succession of conformal diagrams shows from left to right how the surfaces of the  
hourglass foliation are
deformed from nearly light cones to nearly planes as $\tau_{0}$ increases from a very small 
value to a very large one. To better illustrate the effect, different members of  the foliation are
 shown in the different figures of the sequence; but, to keep 
track of the deformation, the surface at $t=0$ (shown with a heavy line) in all cases. 
The Penrose diagram has been doubled to admit negative values of $r$ in the left 
triangular area. This doubling shows how the curves of constant $t,\vartheta,\varphi$ 
are smooth spacelike curves that connect asymptotically past and future null 
infinities.
 The scale of the lenght $\tau_0$ is irrelevant for the effect described in the figure, which only
  depends on the ratios between the different $\tau_0$'s shown.}
\end{center}
\end{figure}

\section{Electromagnetic field in Minkowski space}

\label{sec:EMMink}

We will analyze in this section the case of the electromagnetic field
on a fixed Minkowskian background. Practically all the features that
will be encountered in the gravitational case already appear in this
technically simpler context. 

The main difference which does not hinder the analogy is that, since
the background is fixed, its Poincar\'e symmetry appears as a global
symmetry rather than an asymptotic gauge symmetry. There are no constraints
associated with the surface deformation $\xi$, which are not varied
in the action principle. The $\mH_{\mu}$ in \eqref{Ham} are replaced
by the energy and momentum densities of the electromagnetic field\begin{eqnarray}
\mH_{\perp}^{(elm)} & = & \frac{1}{2}\left(g^{-\frac{1}{2}}\pi_{i}\pi^{i}+\frac{1}{2}g^{\frac{1}{2}}F^{ij}F_{ij}\right)\,,\label{hpe}\\
\mH_{i}^{(elm)} & = & F_{ij}\pi^{j}\,.\label{hie}
\end{eqnarray}
The only gauge symmetry present in the problem is the electromagnetic
one, whose generator is 
\begin{equation}
{\cal G}=-\pi_{,i}^{i}\approx0\,.\label{gg}
\end{equation}
Here $A_{i}$ is the vector potential, $\pi^{i}$ its conjugate momentum,
and $g_{ij}$ is the metric on the hourglass, and $g$ denotes its
determinant.

If instead of having a fixed background we were considering dynamically
coupled electromagnetic and gravitational fields, then expressions
\eqref{hpe}, \eqref{hie} would be added to their gravitational counterparts 
discussed in section \ref{sec:Gravfield}, 
and the sum would be constrained to vanish. The asymptotic analysis given
below would still hold because at large distances the spacetime would be flat. Then the asymptotic
symmetry transformations of the coupled Einstein-Maxwell system would be those discussed
here (internal electromagnetic, and Poincar\'e transformations) and the additional
gravitational supertranslations.

We will now discuss the Poincar\'e and proper and improper gauge transformations
for the electromagnetic field on the hourglass slicing. In this case
the time equal constant surface is left invariant under the Lorentz
group, whereas it is mapped onto a different hyperboloid by spacetime
translations. Thus if one compares the situation with $t=$ constant
planes, one sees that the roles of spatial translations and boosts
are interchanged.

\subsection{Asymptotic boundary conditions}
\subsubsection{Power expansion near $r=\pm \infty$}

By applying the procedure described at the end of Sec. \ref{secII},
starting from the Coulomb field written in hyperbolic coordinates,
one is led to the boundary conditions, 
\begin{eqnarray}
A_{a} & = & a_{a}^{(0)}+ a_{a}^{(1)}r^{-1}+\mathcal{O}(r^{-2})\,,\label{a1}\\
A_{r} & = & a_{r}^{(2)}r^{-2}+\mathcal{O}(r^{-3})\ ,\label{a2}\\
\pi^{a} & = & \pi_{(2)}^{a}r^{-2}+\mathcal{O}(r^{-3})\,,\label{a3}\\
\pi^{r} & = & \pi^r_{(0)}+\mathcal{O}(r^{-1})\, , \label{a4} \ \ \ \ \ \ \ \ \label{a4}\\
\lambda &=& \lambda_{(0)}+\mathcal{O}(r^{-1}) ,\label{gauges}
\end{eqnarray}
 Here $\lambda$ is the Lagrange multiplier that accompanies the gauge generator \eqref{gg}. In addition to the power law decays \eqref{a1}--\eqref{gauges} it is necessary to introduce parity conditions. This is achieved by splitting some of the variables in longitudinal and transverse parts as follows
\begin{eqnarray}
a^{(0)}_{a}=\nabla_{a}F+\star\nabla_{a}\bar{G}\, ,\label{eq:divcurl} \\
h_{a}=\gamma^{\frac{1}{2}}\left(\nabla_{a}N+\star\nabla_{a}\bar{N}\right)\, .\label{eq:divcurl2}
\end{eqnarray}
Here, 
\begin{equation}
\star\nabla_{a}=\gamma^{\frac{1}{2}}\epsilon_{ab}\nabla^{b}\,\label{nablabar},
\end{equation}
where $\gamma$ is the determinant of the metric $\gamma_{ab}$ on the unit two-sphere. The ``news'' vector $h^a$ in \eqref{eq:divcurl2}, which will play a central role in what follows, is defined by
\begin{equation}
h^{a}=\frac{1}{\tau_{0}^2}\left(\pi^{a}_{(2)}+\gamma^{\frac{1}{2}}\gamma^{ab}f_{b r}^{(2)}\right)\, ,\label{ha}
\end{equation}
for $r\rightarrow\pm\infty$, where the
 $f^{(2)}_{br}$ is the leading order coefficient of $F_{br}$.
In Minkowski coordinates the news correspond to an electromagnetic
field that decays as $r^{-1}$, that is to a wave emerging from a confined source $(r\rightarrow+\infty)$, or converging towards an absorber$(r\rightarrow-\infty)$. For an accelerating electric charge $e$ one has from the Lienard-Wiechert field, 
$$
\gamma^{\frac{1}{2}}f^{ar}_{(2)}\vec{\partial}_{a}=\pi_{\left(2\right)}^{a}\vec{\partial}_{a}=-2e\gamma^{\frac{1}{2}}\hat{r}\times\left(\hat{r}\times\vec{a}\right),
$$
where $\vec{a}$ is the acceleration
 in rest frame of the emitter (outgoing wave) or absorber (incoming wave).
See for example, \cit{Rohrlich,Teitelboim:1970mw}. 

  On the entire hourglass this field stems from
$$
F_{rad}=F_{ret} - F_{adv}
$$  
introduced by Dirac\cit{Dirac:1938nz}, which is derived from the function, 
$$
\frac{1}{r}\delta(t-r),  \ \ \ \ \ -\infty < r< \infty \ .
$$
This function is not a Green function, but a solution of the homogeneous equation: the difference $F_{ret} - F_{adv}$. Therefore it has as much radiation coming in as going out, and obeys the parity conditions that will be discussed next.

\subsubsection{Parity conditions}

The parity conditions will be then the following,
\be{parity}
\left. (F,\lambda_{(0)}, N,\bar N)\right|_{+\infty} =
\left. (F,\lambda_{(0)}, N,\bar N)\right|_{-\infty} \ \ 
\mbox{for each } (\vartheta,\varphi).
\ee
 
 Parity conditions play a fundamental role in the Regge-Teitelboim discussion of Poincar\'e invariance
on asymptotic planes. We see that when dealing with Bondi, Metzner, Sachs invariance on hyperboloids,
they again come in\footnote{The BMS symmetry has been tamed to fit a foliation by surfaces that are
asymptotically planes  \cit{Henneaux:2018cst,Henneaux:2018gfi,Henneaux:2018hdj,Henneaux:2019yax}.
This has required dexterity, since the symmetry is intimately related to radiation and
its natural habitat is an asymptotically null surface, rather than a plane.}. 
The reason for their appearance  is discussed next.

\subsubsection{Surface deformation algebra}
\label{sdfalgebra}

The surface deformations involved in the present analysis are Poincar\'e deformations of the
 hypersurface, which are motions generated by the Killing vectors given in appendix \ref{app1}, and besides them, gauge transformations. If one evaluates the commutator of any two
 such deformations one finds the following results for the asymptotic part of the 
 commutator:

 (i) Two Poincar\'e deformations close according to the Poincar\'e 
 group. 
 
 (ii) Spacetime translations commute with improper gauge transformations. 
 
 (iii) The commutator of a Lorentz transformation with Killing vector 
 $\xi_{\text{Lorentz}}$ and an improper gauge transformation with parameter $\lambda(\vartheta,\varphi)$ 
 at infinity, is a gauge transformation with parameter, 
 \be{et}
 \zeta=\xi_{\text{Lorentz}}^{a}\partial_{a}\lambda.
 \ee
 
 The results (i)--(iii) are the expressions, in terms of deformations, of the 
 electromagnetic BMS algebra.
 
 Equation \eqref{et}, which technically stand from the fact that the Lorentz group acts asymptotically 
 as a Lie derivative on the two sphere (appendix \ref{app1}) has a profound 
 consequence:  it mixes improper gauge transformations with spacetime 
 motions. This provides a tantalizing confirmation of the fact that improper 
 gauge transformations are physically relevant motions that cannot be ``factored 
 away''. 
 
\subsection{The hyperbolic hourglass as an unconventional Cauchy surface}

The physical motivation for the parity conditions is very simple. They essentially state that for a
closed system (the free electromagnetic field in this case) everything that comes
in must come out. That is, one allows for non--vanishing incoming and outgoing fluxes of energy,
momentum, and other (BMS) charges; but requires that the net flux should be equal to zero.

This requirement, which physically is a condition connecting the remote past with the remote future,
can be formulated as a fixed time statement, because the spacelike hyperbolic hourglass is
asymptotically tangent to the past and future lightcones. This is the reason 
for bringing in the hourglass in the first place.

When regarded as an initial value surface, the hourglass has the unconventional 
feature, that a point, which is not at infinity, lying, say, on the outgoing half of
 one hourglass at a given time, also lies on the incoming half of another hourglass at a later time. 
 This implies that one cannot give freely initial value data on the complete hourglass but only on half of it,
 the outgoing half for example. However, the double counting of points does not happen at infinity, so if one 
 gives data on the outgoing half one should specify additionally the incoming radiation, that is one should give
 the news at $r=-\infty$. But this is precisely what the parity condition does, stating that the 
 incoming news are equal to the outgoing ones. Thus it is sufficient to specify just the data on the outgoing
  half of the hourglass (or, viceversa, on the incoming one) if the parity condition is imposed.
  
Therefore one must bring in the complete hourglass in order to deal in Hamiltonian terms with the 
interrelationship
 between past and future, but one only gives initial value data on one half of it, together with asymptotic 
 information on the other half. In this sense the hourglass  is a Cauchy surface.

\subsection{Poincar\'e invariance of the boundary conditions. Fiber memory}
\label{sec:Poinc}

One must demand, by consistency, that the boundary conditions \eqref{a1}--\eqref{gauges} and the parity conditions \eqref{parity} should be preserved  under the Poincar\'e group. This can be efficiently analyzed by writing the equations of motion in Hamiltonian form, 
\begin{eqnarray}
\dot{A}_{i} & = & -F_{ij}\xi^{j}+g^{-1/2}\xi^{\perp}\pi_{i} +\partial_{i}\lambda\,,\label{dA-1}\\
\dot{\pi}^{i} & =
 & (\pi^{i}\xi^{j})_{,j}-\pi^{j}\xi_{,j}^{i}+(\xi^{\perp}F^{ji}g^{1/2})_{,j}\ ,\label{dpi-1}
\end{eqnarray}
and taking as the deformation parameters that multiply the generators \eqref{hpe}, \eqref{hie} to be the components, $\xi^\perp$ and $\xi^i$  of the Poincar\'e Killing vectors  given in appendix \ref{app1}.

 The preservation of the boundary conditions \eqref{a1}--\eqref{gauges}  is straightforward. One first verifies Lorentz invariance which is simpler because the hyperbolic hourglass is Lorentz invariant. With this established, it is sufficient to check invariance under time translations, which is more laborious but straightforward.
 
 Next one turns to the conditions \eqref{parity}. The preservation of the parity conditions under Lorentz transformations is evident because they map each asymptotic region into itself. One needs then only be concerned with time translations. Under them, the equation of motion for the leading order term  of $A_a$ is,
 \be{dotAa}
 \dot a_a^{0}=\gamma^{-\frac{1}{2}}h_a,
\ee
 for $r\rightarrow\pm\infty$. Its longitudinal component is
\be{Fdott}
\dot F =N,
\ee
which shows that the preservation in time of the parity of $F$ is automatically satisfied because of the parity conditions on $N$. The parity of $F$ is also preserved under improper gauge transformations because of the parity condition on $\lambda_{(0)}$. For $\lambda_{(0)}$ itself there is nothing to check because it has no equation of motion.  

The preservation of the parity conditions for $N$ and $\bar N$ pose no problem either, as it is shown in appendix \ref{app2}.

Equation \eqref{Fdott} has a highly non trivial content.
It shows that, even when the generator of improper gauge transformations does not act, i.e., when  $\lambda_{(0)}=0$, and one is only moving in the time $t$, there is still a displacement,
\be{Ndt}
\delta F=N\delta t 
\ee
along the U(1) fiber of amount $N\delta t$ when a time $\delta t$ elapses. That is: (i) If there are no news (and one does not change the gauge frame) $F$ is conserved, (ii) If there are news during a time interval the value of $F$ changes from $F_{\mbox{\tiny before}}$ to $F_{\mbox{\tiny after}}$ according to the integral of  \eqref{Fdott} over the time interval. That is, $F$ ``remembers'' the news, and for that reason is called the ``fiber memory''. Another kind of memory, ``charge memory'' will be encountered below in section \ref{ssrates}.

It is important to realize that \eqref{Ndt} is not just a ``redefinition of $\lambda_{(0)}$ by the amount $N$". This is because $\lambda_{(0)}$ is not in the phase space, and can be held fixed in the variation of the Hamiltonian, whereas $N$ is a dynamical variable, which obeys a (gauge invariant) equation of motion and hence cannot be held fixed.

\subsection{BMS charges}

\subsubsection{Electric BMS charge}

Taking into account the parity condition on $\lambda_{(0)}$ one finds that the surface integral that must be added to the electromagnetic gauge generator to include improper transformations is given by
\be{gaugeimp}
\oint \lambda_{(0)} Q
\ee 
where the gauge charge $Q$ is given by
\begin{equation}
Q\left(\vartheta,\varphi\right)=\left.\pi_{(0)}^{r}\right|_{+\infty}-\left.\pi_{(0)}^{r}\right|_{-\infty} \equiv Q_+ + Q_-.\label{eq:Qgauge}
\end{equation}
It is important to interpret this expression appropriately. The hourglass is a construct that enables one to keep track, within the Hamiltonian formalism, of the incoming and outgoing radiation in an economic manner, that is  without introducing separate overlapping incoming and outgoing hyperbolic patches. This brings in a redundancy: one way or another space its counted twice. We just saw one instance of this above in connection with the initial value data. The redundancy strikes again in expression \eqref{eq:Qgauge} for the charge. If one considers the Coulomb field of a particle of charge $e$ at rest at $x^i=0$ one finds,
$$
Q_+\left(\vartheta,\varphi\right) =\frac{e}{4\pi}\sin\vartheta
$$
and
$$
Q_- \left(\vartheta,\varphi\right)=\frac{e}{4\pi}\sin\vartheta
$$
and hence
\be{Qcoul}
Q\left(\vartheta,\varphi\right)= 2\frac{e}{4\pi}\sin\vartheta
\ee
The factor two arises because one is counting twice: $Q_+$ is the charge as seen in the
 outgoing description of space, while $Q_-$ is the {\it same} charge as seen from its incoming replica.
  This point will reappear below in connection with radiation rates.

\subsubsection{Magnetic BMS charge}

There is a magnetic analog of \eqref{eq:Qgauge} given by 
\be{barQ}
\bar Q=\left.\epsilon^{ab}\nabla_a a^{(0)}_{b} \right|^{+\infty}_{-\infty}=\left.\gamma^{1/2}\nabla^2\bar G\right|^{+\infty}_{-\infty} ,
\ee
which is conserved as a consequence of \eqref{dotAa} and the parity condition 
for $\bar{N}$, 
 \be{dQ0}
 \dot{\bar Q}=0,
 \ee

 In the electric representation this conservation law  appears as an ``accidental'', because it does
  not follow from a symmetry of the action. The formalism becomes complete if one introduces
   a second potential, so that the electric and magnetic charges are treated on the same footing.
    This completion of the formalism may be regarded as a matter of elegance and economy, but not
     of necessity, for questions that can be asked within the electric representation. But, as we will see
     further below, it becomes essential when one discusses Lorentz transformations. Therefore we recall it right
       away.

\subsection{Asymptotic two potential formulation}

\label{mm}

One brings in a new, 
``magnetic" vector potential $\bar{A}$. For the present purposes it is sufficient to do so only asymptotically.  The potential $\bar{A}$ satisfies, 
\be{abardef}
\pi^{r}=-\epsilon^{ab}\partial_{a}\bar{A}_{b}\,,\ \ \ \ \pi^{a}=-\epsilon^{ab}\left(\partial_{b}\bar{A}_{r}-\partial_{r}\bar{A}_{b}\right)\,.
\ee
Then, equations \eqref{a3}, \eqref{a4} are replaced by 
\begin{eqnarray}
\bar{A}_{a} & = & \bar{F}_{,a}-\gamma^{-\frac{1}{2}}\gamma_{ac}\epsilon^{cb}G_{,b}+\bar{a}_{a}^{(1)}r^{-1}+\mathcal{O}(r^{-2})\,,\label{bcm1}\\
\bar{A}_{r} & = & \bar{a}_{r}^{(2)}r^{-2}+\mathcal{O}(r^{-3})\,.\label{bcm2}
\end{eqnarray}

It is important to realize that the new potential $\bar{A}$ incorporates with it the \textit{additional} 
variable $\bar{F}$, which was not present in the electric representation and drops out 
from eqs. \eqref{abardef}.

 There are now also magnetic
gauge transformations with an associated parameter $\bar{\lambda}_{\left(0\right)}$,
which is independent of the ``electric'' $\lambda_{\left(0\right)}$.
Under a magnetic BMS transformation $\bar{F}$ and $G$ transform
according to
\begin{equation}
\bar{F}\rightarrow\bar{F}+\bar{\lambda}_{\left(0\right)}\,,\label{eq:F-1}
\end{equation}
\begin{equation}
G\rightarrow G\,.\label{eq:G-1}
\end{equation}
Here the electric and magnetic radial momenta $\pi^r$, $\bar\pi^r$, are related  $G$ and $\bar G$ through,
\be{piG}
\pi^r= \gamma^{\frac{1}{2}}\nabla^2 G, \ \ \ \ \ \ \bar\pi^r= \gamma^{\frac{1}{2}}\nabla^2 \bar G.
\ee

If one demands that $G$ and $\bar{G}$ be regular on the sphere,
there is no room for a zero mode in
the electric and magnetic BMS charges.  The zero modes must be introduced through Dirac string singularities. 

For a magnetic pole of strength $g$ at the origin, on has 
\begin{equation}
A_{\varphi}=g(1-\cos\vartheta)\,,\label{Apole}
\end{equation}
\begin{equation}
\bar{G}=g\log(1+\cos\vartheta)\, .\label{Gpole}
\end{equation}
For an electric pole of strength $e$, which in the electric representation has
\begin{equation}
\pi_{(0)}^{r}=\gamma^{\frac{1}{2}}e\,,\label{qpole}
\end{equation}
one now writes 
\begin{equation}
\bar A_{\varphi}=e(1-\cos\vartheta)\,,\label{epole}
\end{equation}
\begin{equation}
G=e\log(1+\cos\vartheta)\, .\label{egpole}
\end{equation}
If one admits Dirac string singularities in $G$ and $\bar G$ one must also do so for $F$ and
 $\bar{F}$ in order, for example, to be able to implement rotations. This is so because under
  a rotation the monopole potentials change by a singular gauge transformation.
  
  \subsubsection{Electric-magnetic duality invariant notation}
  
  It is useful to introduce a compact notation that makes electric-magnetic duality
   invariance of the theory manifest. This is achieved by writing 
   \be{AM}
A_a^M=\partial_a F^M +\epsilon^{M}_{\ \ N}\star\nabla_a G^N,
\ee
\be{twopotnot}
 A^{M}=\begin{pmatrix}A\\
\bar{A}
\end{pmatrix}, \ \ \
 N^{M}=\begin{pmatrix}N\\
\bar{N}
\end{pmatrix}, \ \ \ F^{M}=\begin{pmatrix}F\\
\bar{F}
\end{pmatrix},  \ \ \  G^{M}=\begin{pmatrix}G\\
\bar{G}
\end{pmatrix}. 
\ee
where
\be{qm}
Q^M=\left.\frac{}{}
\gamma^{\frac{1}{2}}\nabla^2 G^M \right|^{+\infty}_{-\infty},
\ee
are the electric and magnetic charges.  
  
\subsection{Spacetime translations: Improved generator}

\label{imp}

\subsubsection{Analysis starting from the electric representation}

Rather than employing the electric-magnetic invariant formalism ab initio, we prefer to start from the ``electric" representation and then use elements of duality to ``patch it" in order to cast final results in a duality invariant form. This we do for expediency, but -- more importantly -- because in the case of gravitation, where the full duality invariant formalism has not yet been developed, one can still perform the same steps, starting from the available electric representation.

If one considers the Hamiltonian for a motion corresponding to a time 
translation, the surface term in the variation of the 
Hamiltonian 
\begin{equation}
H_{0}\left[\xi\right]=\int d^{3}x\left(\xi^{\perp}\mH_{\perp}^{(elm)}+\xi^{i}\mH_{i}^{(elm)}\right)\,,\label{H0}
\end{equation}
is given, in the electric representation, by 
\be{dHzero}
\delta H_0=\left. \oint\left(\alpha^{\mu}k_{\mu}\right)h^{a}\delta a_{a}^{\left(0\right)}\right|^{\infty}_{-\infty}  \,
\ee
where $\alpha^{\mu}$ is the amount of spacetime translation and 
\begin{equation}
k_{\mu}=\left(-1,\hat{r}\right)\,.\label{eq:kmuret}
\end{equation}
Equation \eqref{dHzero} may be rewritten separating the electric memory and magnetic charge variations as,
\begin{align}
\delta H_{0} =&-\oint\left.\partial_{a}\left[\left(\alpha_{\mu}k^{\mu}\right)\left(\gamma^{\frac{1}{2}}\gamma^{ab}N_{,b}+\epsilon^{ab}\bar{N}_{,b}\right)\right]\delta F\right|_{-\infty}^{+\infty} \nonumber \\
& -\oint\left.\partial_{a}\left[\left(\alpha_{\mu}k^{\mu}\right)\left(\gamma^{\frac{1}{2}}\gamma^{ab}\bar{N}_{,b}-\epsilon^{ab}N_{,b}\right)\right]\delta\bar{G}\right|_{-\infty}^{+\infty}, 
\label{dHO}
\end{align}

If  the parity conditions are used, the first term on the right hand side on \eqref{dHO} vanishes, 
and the second term may be written as 
\begin{equation}
\delta H_0=
\oint \bar{\lambda}_{\text{trans}} \delta \bar{Q},
\label{deltaH}
\end{equation}
where $\bar{\lambda}$ is given by 
\be{lambdabar}
\nabla^{2}\bar{\lambda}_{\text{trans}}=-\partial_{a}\left[\left(\alpha_{\mu}k^{\mu}\right)\left(\gamma^{\frac{1}{2}}\gamma^{ab}\bar{N}_{,b}-\epsilon^{ab}N_{,b}\right)\right].
\ee
Equation \eqref{deltaH} shows that in order to improve $H_0$ one must add to it 
a term proportional to the \textit{magnetic} gauge constraint\footnote{The magnetic gauge generator, $-\bar\pi^i_{,i}$, can be treated properly by keeping in Dirac's ``total Hamiltonian" the full constraint $\vec\pi_{mag}=0$ and $\vec\pi_{el}+\nabla\times\vec {\bar A}=0$, whose curl is second class, while their divergence $\nabla\cdot\vec\pi_{mag}$ is first class. The details of that treatment will not be needed herein.},
\be{impterm}
\int-\bar{\pi}_{,i}^{i}\bar{\lambda}_{\text{trans}}d^{3}x.
\ee
This shows that it is essential to bring in the magnetic sector in order to properly define the spacetime 
translation generators. \textit{The improvement cannot be made solely within the 
electric sector}. In other words, a deformation consisting only of a spacetime translation by itself  
does not have a well-defined generator. Only when one adds to it a movement 
along the fiber whose magnitude is $\bar{\lambda}$ given by \eqref{lambdabar}, thus the generator exists.
 It is this improved generator which deserves to be called 
 $\alpha^{\mu}P_{\mu}$. The numerical value of $P_{\mu}$ is the same as the 
 original $H_{0}$ because the other term \eqref{impterm} vanishes weakly\footnote{One could have
  try to stay within the electric sector by demanding that the magnetic charge $\bar{Q}$ should be a passive espectator given as an
  ``external field'', and not varied in the action principle. For consistency it should be given so that $\dot{\bar{Q}}=0$
  (eq. \eqref{dQ0}) up to a Lorentz transformation. But the boundary term in \eqref{dHO} would not vanish if $\delta \bar G=\bar G_{,a}\xi^a$, so this possibility is not tenable if one wants to have Lorentz invariance. Thus it is Lorentz invariance which forces one to bring in the magnetic sector with its own independent life.}
   
  \subsubsection{Simplification for time translations. Magnetic fiber memory brought in}
  
  For the case of the time translations, for which 
\be{alpha0}
-\alpha_{\mu}k^{\mu}=\alpha^0,
 \ee  
 two simplifications occur that are worth 
  noting and will be useful later on:
  
  (i) The term proportional to $\epsilon^{ab}\delta F$ in \eqref{dHO} vanishes, 
    
  (ii) The compensating magnetic gauge transformation $\bar{\lambda}$ reads just  
\be{barlambdat}
\bar{\lambda}=\bar{N}. \qquad \qquad (\text{time translation})
\ee
Equation \eqref{barlambdat} permits to understand the need for the addition of the magnetic gauge transformation. {\it It simply brings in the magnetic fiber memory}, that -- unlike the magnetic charge -- is not present in the purely electric formulation, because only the gauge invariant curl of the magnetic potential appears in it.

\subsubsection{Full implementation of duality}

The above discussion cannot be yet complete because after just including \eqref{impterm} the variation of the Hamiltonian would read
\be{H0dinv}
\delta H_{0}=\left.-\oint\partial_{a}\left[\left(\alpha_{\mu}k^{\mu}\right)\left(\gamma^{\frac{1}{2}}\gamma^{ab}N_{,b}+\epsilon^{ab}\bar N_{,b}\right)\right]\delta F\right|_{-\infty}^{+\infty},
\ee
and this expression is not duality invariant. Had we started from the magnetic representation instead, we would have obtained
\be{H0dinvmag}
\delta H_{0}=\left.-\oint\partial_{a}\left[\left(\alpha_{\mu}k^{\mu}\right)\left(\gamma^{\frac{1}{2}}\gamma^{ab}\bar N_{,b}-\epsilon^{ab} N_{,b}\right)\right]\delta \bar F\right|_{-\infty}^{+\infty},
\ee
and $\lambda_{\text{trans}}$ would have been given by
\be{lambdabarmag}
\nabla^{2}{\lambda_{\text{trans}}}=-\partial_{a}\left[\left(\alpha_{\mu}k^{\mu}\right)\left(\gamma^{\frac{1}{2}}\gamma^{ab}{N}_{,b}+\epsilon^{ab}\bar N_{,b}\right)\right].
\ee
in order to bring in the electric memory.

In \eqref{H0dinv} and \eqref{H0dinvmag} above, the equality means that terms proportional to the constraints, $-\pi^i_i$ and $-\bar \pi^i_i$ have been dropped.

It is evident, just from demanding duality invariance, that the correct result, which incorporates \eqref{H0dinv} and \eqref{H0dinvmag} should read,
\be{H0dinvdual}
\delta H_{0}=\left.-\oint\partial_{a}\left[\left(\alpha_{\mu}k^{\mu}\right)\left(\gamma^{\frac{1}{2}}\gamma^{ab}N_{,b}^{M}+\epsilon^{MN}\epsilon^{ab}N_{N,b}\right)\right]\delta F_M\right|_{-\infty}^{+\infty},
\ee
which for an infinitesimal time translation of magnitud $\alpha^0$ simplifies to
\be{H0dinvdual2}
\delta H_{0}=-\left.\alpha^0 \oint\gamma^{\frac{1}{2}}\nabla^a N^{M}\delta (\nabla_a F_M)\right|_{-\infty}^{+\infty}.
\ee

\subsection{Lorentz generators. Spin from charge}

We again start from the electric representation and at the end cast the results in a manifestly duality invariant form.

We have
\be{Hlorentz}
H_0^{\mbox{\tiny Lorentz}}=-\int d^3 x \xi^i F_{ij}\pi^j,
\ee
where $\xi$ are the Lorentz Killing vectors.
The surface term in its variation reads
\be{surflor}
\delta H_0^{\mbox{\tiny Lorentz}}=-\oint \left. \xi^{a}\delta 
A_{a}\pi^{r}\right|^{+\infty}_{-\infty}.
\ee
To improve the generator $\mH_0$ we add an electric gauge generator,
but this time \textit{with the surface term included}, namely, 
\be{GLorentz}
\mathcal{G}^{\mbox{\tiny Lorentz}}=\left.\int-\pi_{,i}^{i}\lambda_{\mbox{\tiny Lorentz}}\,d^{3}x+\oint\lambda_{\mbox{\tiny Lorentz}}\left(\infty\right)\pi^{r}\right|_{-\infty}^{+\infty},
\ee
with 
\be{llorentz2}
\lambda_{\mbox{\tiny Lorentz}}(\infty)=\xi^{a}_{\mbox{\tiny Lorentz}}A_a.
\ee
  One then finds that the variation of 
 \be{Himplor}
{H}_{\mbox{\tiny improved}}^{\mbox{\tiny Lorentz}}= H_0^{\mbox{\tiny Lorentz}}+\mathcal{G}^{\mbox{\tiny 
Lorentz}},
 \ee
 does not have a surface integral. 
 
 \subsubsection{Lie derivative restored}
 
 The improvement of the Lorentz generator $\mH_0^{\mbox{\tiny Lorentz}}$ has an important 
 geometrical consequence, in that it restores the Lie derivative at infinity. Indeed, the change in
  $A_{i}^{M}$ given by the generator $\mH_0^{\mbox{\tiny Lorentz}}$ is given by 
 \[
 \delta_{0}A_{i}=\xi^{j}F_{ji}=\mathcal{L}_{\xi}A_{i}-\partial_{i}\left(\xi^{j}A_{j}\right),
 \]
 so that 
 \[
 \delta_{\mbox{\tiny 
 improved}}A_{a}\left(\infty\right)=\mathcal{L}_{\xi}A_{a}\left(\infty\right).
 \]
 Therefore, $\mathcal{H}_{\mbox{\tiny improved}}^{\mbox{\tiny Lorentz}}$ is the 
 generator that will correctly implement the deformation algebra in section 
 \ref{sdfalgebra}.
 
 \subsubsection{Spin from charge}
 
 The numerical value of the generator \eqref{Himplor} which realizes the improvement of the 
 Lorentz generator is not zero, but it is equal to the surface integral that 
 appears in it. Therefore the numerical value of the angular momentum is not 
 just the volume integral \eqref{Hlorentz}, but it includes a contribution 
\be{lorimpgen}
\left.\oint \xi^{a} A_{a}\pi^{r}\right|_{-\infty}^{+\infty}= \left.\oint \pi^{r} \xi^{a} \partial_{a}F\right|_{-\infty}^{+\infty} + S,
\ee 
where
 \be{stt}
S = \left.\oint  \xi^{a}\star\nabla_a \bar G \gamma^{\frac{1}{2}}\nabla^2 G \right|_{-\infty}^{+\infty}=-\frac{1}{2}\left.\oint  \xi^{a}\star\nabla_a G^M\epsilon_{MN} \gamma^{\frac{1}{2}}\nabla^2 G^N \right|_{-\infty}^{+\infty},
 \ee
which is proportional to the electric BMS charge 
$\pi^{r}=\gamma^{1/2}\nabla^2 G$. This phenomenon is similar
to the modification of the angular momentum which appears in the presence
of a magnetic pole in abelian and non-abelian gauge theories.The
novelty here is that it occurs already without a magnetic pole.

The spin from charge phenomenon does not happen for energy and momentum
because no surface term analogous to the one appearing in \eqref{lorimpgen}
is included in the translation charge.

\subsubsection{Duality invariant Lorentz generator}
\label{dilg}

The improved electric Lorentz generator,
\be{electricJ}
H_{el}[\xi]=\left. H_0[\xi]+S+ \oint \pi^r \xi^a\partial_a F\right|_{-\infty}^{+\infty},
\ee
Is not electric-magnetic duality invariant because, whereas $H_0$ and $S$ have that property, the term proportional to $\pi^r$ does not. Just as it was discussed for translations, it is evident that the appropriate expression is
\begin{eqnarray}
	 H[\xi]&=& \left. H_{el}[\xi] + \oint \bar\pi^r \xi^a\partial_a \bar F\right|_{-\infty}^{+\infty}
=\left. H_0[\xi]+S+ \oint \pi^r_M \xi^a\partial_a F^M\right|_{-\infty}^{+\infty}\nonumber \\
&=& H_0[\xi]+S+ \oint Q_M \xi^a\partial_a F^M. \label{lorentzQ}
\end{eqnarray}
One may think of $\oint\xi^a Q^M F_{M,a}$ as the generator of Lorentz transformations at infinity, and $H_0+S$ as the ``bulk part" (although $S$ is a surface integral).

It will be shown below that the duality invariant angular momentum is conserved (the electric part \eqref{electricJ} is not!).  Since this has been an issue in the literature (in the case of gravitation, which will follow the same lines) it is worth some comment.

First of all one realizes that under improper electric and magnetic gauge transformation, with parameter $\lambda^M_{(0)}=\epsilon^M$, the  Lorentz generator changes as,
\be{Lortran}
H[\xi] \longrightarrow H[\xi]  - \oint \epsilon^M \partial_a(\xi^a Q_M),
\ee 
and the new angular momentum is also conserved.

This is just as it happens if one changes the origin for orbital angular momentum, and in our view it is not to be regarded as a difficulty, since the present formalism improper gauge transformations are on the same footing with spacetime translations. All the more so, since a ``pure time translation" carries along with it a rotation along the fiber, due to the fiber memory.

\subsection{Symmetry algebra}

Applying the general formula \eqref{pbH} to the present case one obtains that the Poincar\'e generators obey the Poincar\'e algebra,
the electromagnetic charges are abelian. Furthermore, the electric and magnetic charges $Q_M(\vartheta,\varphi)$ transform as 
\begin{equation}
\left[Q_M,\mH_{\mbox{\tiny Lorentz}}\right]^*=\ \partial_{a}\left(Q_M\xi_{\mbox{\tiny Lorentz}}^{a}\right)\,,\label{QHcom}
\end{equation}
under Lorentz transformations and they obey
\begin{equation}
\left[Q_M\left(\vartheta,\varphi\right),P_{\mu}\right]^*=0\, ,\label{eq:QP}
\end{equation}
so that they are invariant under spacetime translations.

Equation \eqref{QHcom} may be verified by inspection from \eqref{lorentzQ} because under a gauge transformation $F_M\rightarrow F_M+ \epsilon_M$, $\delta J=\oint Q^M\epsilon_{,a}\xi^a=-\partial_a(Q^M\xi^a)$, which is what \eqref{QHcom} yields, read as the action of $Q_M$ on $\vec J$. Similarly for the boosts.

Equations \eqref{QHcom} and \eqref{eq:QP} have been written in term of Dirac brackets because the terms proportional to gauge constraints that accompany the surface integrals to make a well defined generator have been dropped.

\subsection{Emission and absorption rates. Charge memory} 
\label{ssrates}

\subsubsection{General formula for emission rates}

Our boundary conditions are appropriate for a closed system, whose Hamiltonian is invariant
 under Poincar\'e and improper gauge transformations, and the corresponding conservation laws hold as a consequence of the fact that as much radiation is coming in as going out.
 
However, the formalism provides expressions for the emission and absorption rates separately. For that purpose one realizes that
\begin{equation}
\dot Q_\alpha =
-\oint h^{a}\delta_{\alpha}a_{a}^{\left(0\right)}\bigg|^{+\infty}_{-\infty} \,
=\left.-\oint \gamma^{\frac{1}{2}}\nabla^a N^M \nabla_a(\delta_{\alpha}F_M)\right|^{+\infty}_{-\infty} .\label{rates} 
\end{equation}
Here $\delta_{\alpha}F^M$ is the change of $F^M$ generated by 
the charge $Q_\alpha$.
Thus  $\delta F^M=\epsilon^M$
for gauge transformations, $\delta F^M=N^M$
for time translations, and $\delta F^M= F^M_{\ \  ,a}\xi^a$
for rotations. The purely electric form does not exist for the Lorentz charges.  

Then, the emission rates are read from the upper endpoint in \eqref{rates} and the absorption rates from the lower one. In this way, one obtains the following results.

\subsubsection{BMS charge}

\be{nnnn}
\dot Q^M=\partial_{a} h^{a}_{M}\Big|^{+\infty}_{-\infty} =\gamma^{\frac{1}{2}}\nabla^2 N^M \Big|^{+\infty}_{-\infty},
\ee
where,
\[
h_{N}^{a}=\begin{pmatrix}h^{a}\\
-\star h^{a}
\end{pmatrix}.
\]
This equation is to be interpreted as giving either $\left[+\dot\pi^r (\infty)\right]$,
 or $\left[-\dot\pi^r (-\infty)\right]$. These are {\it not} to be thought of as the rate of change of
  two different charges, but rather as the rates of change of one and the same charge, due to
   outgoing and incoming radiation respectively; which must be calculated using the two replicas
    of space that form the hourglass. When the parity conditions hold the $Q^M$ are conserved.

On sees from \eqref{nnnn}, in analogy with \eqref{Fdott}, that the BMS charge
 also ``remembers'' the news and that, in this sense, the Laplacian of $N$ is the ``charge memory''.
 We will see in [\citenum{BGP2}], that when a cosmological constant is introduced the fiber and charge memories are
 different and that the fiber memory appears to be more fundamental.

\subsubsection{Energy}

Similarly, one finds for the energy 
\begin{eqnarray}
\frac{dP^{0}}{dt} &=&-\left.\oint \gamma^{-\frac{1}{2}}\gamma_{ab}h^{a}h^{b} \right|^{\infty}_{-\infty} \nonumber
\\ &=& \left.-\oint \gamma^{-\frac{1}{2}}\nabla^a N^M \nabla_a N_M\right|^{\infty}_{-\infty} .\label{dtP} 
\end{eqnarray}

\subsubsection{Angular momentum}
\label{angmom}
The equations for the rate of
 change of the BMS charges and the energy given above can be
 expressed solely in terms of quantities
  defined in the electric sector. This is not the case for the angular momentum which as argued before,
   needs the magnetic sector for its very definition. Therefore,  the rate can be read only from the second
 term on the right hand side of  eq. \eqref{rates} , which yields,  
\begin{equation}
\frac{d\vec{J}}{dt}=\left.-\oint \gamma^{\frac{1}{2}} \nabla^a N^M\nabla_a\left(\L_{\vec{\xi}} F_{M}\right) \right|^{\infty}_{-\infty}\,\label{dtJ},
\end{equation}
an expression that can be rewritten as,
\begin{equation}
\frac{d\vec{J}}{dt}=\oint \dot Q_M  F^M_{\ ,a}\vec \xi^a.\label{dtJQ}
\end{equation}

The last expression shows that when the parity conditions hold, so that $\dot Q^M=0$, the angular momentum, Eq. \eqref{lorentzQ}, is conserved, as it was announced and discussed in Sec \ref{dilg}.

 Note that eq. \eqref{dtJ} involves the 
   variable $\bar F$ which does not appear in the electric sector. This is a consequence, in turn, 
   of the fact that the angular momentum changes under the action of the magnetic BMS charge.

The interpretation of these equations is that the left hand sides are the rate of change of one and the
 same energy and angular momentum due to outgoing and incoming radiation. Therefore, the volume 
 integrals appearing in the definition of $P_0$ and $J$ (see eq. \eqref{H0}), are to be thought of as
  evaluated on the upper half of the hourglass in the calculation of outgoing radiation and on the lower
   half in the calculation of incoming radiation. One does not integrate over the whole hourglass
    because this would lead to the same overcounting  encountered for the electromagnetic 
    charges. 
    
    Just as it was the case with the angular momentum itself, the physical cogency of Eq. \eqref{dtJ} giving its rate of change, deserves a brief comment. The time rate of change of $F^M$ is invariant under (improper) gauge transformations. If one agrees to keep the gauge frame fixed, that is, if one only moves in the course of time on the fiber as dictated by the fiber memory, then $F_M(t)$ is determined by the equations of motion -- in a gauge invariant manner - once $F_M(t=0)$ is given. This means that if one were absorbing angular momentum at infinity so as to, say, make a top start spinning, then one would in principle be able to determine $F_M(t=0)$ and thus learn how the BMS origin in \eqref{dtJ} is shifted from the one arbitrarily chosen on the fiber.

\section{Gravitational field}

\label{sec:Gravfield}

\subsection{Correspondence with electromagnetism}

\label{subsec:corresp}

In this section we analyze the gravitational field along the same
lines that we analyzed above the electromagnetic field. The parallel
between both cases is so close that it permits to make the following
discussion succinct. The correspondence is as follows: The $\ell=0$
mode of the improper gauge symmetry generated by the total electric
charge $Q$ is the analog of the $\ell=0$, $\ell=1$ modes of the
Bondi-van der Burg-Metzner-Sachs supertranslation, which are the ordinary
translations generated by $P_{\mu}$. The modes with $\ell\geq1$
of the improper gauge symmetry correspond to the modes $\ell\geq2$
of the supertranslations. Therefore, altogether, one has the correspondence:
\[
\underbrace{\left(Q\left(\vartheta,\varphi\right),P_{\mu}\right)}_{\text{electromagnetism}}\longleftrightarrow\underbrace{{\cal P}\left(\vartheta,\varphi\right)}_{\text{gravitation}}\,.
\]
On the other hand, the Lorentz transformations play along side: 
\[
\underbrace{J_{\mu\nu}}_{\text{electromagnetism}}\longleftrightarrow\underbrace{J_{\mu\nu}}_{\text{gravitation}}\,.
\]
There is, as emphasized before, the difference that in the gravitational
case all the generators are given by surface integrals, whereas in
the electromagnetic one since the background was fixed, the spacetime
translations and the Lorentz transformations were not. But this is
just a technical point which is easily accounted for and does not
hinder at all the close correspondence between both cases. 

The important concept of ``news'' is also present here, of course
since it is the context in which it was originally introduced by Bondi
 \cit{Bondi:1960jsa}. The only difference is that now it is a symmetric
traceless tensor $h^{ab}$, appropriate to describe a gravitational
wave, rather than the vector $h^{a}$ appropriate for an electromagnetic
one. Thus, one has the correspondence:
\[
\underbrace{h_{a}}_{\text{electromagnetism}}\longleftrightarrow\underbrace{h_{ab}}_{\text{gravitation}}\,.
\]
Keeping this in mind, we will essentially write the corresponding
equation without much discussion, because one may translate to gravitation
word by word in each case the corresponding comments from electromagnetism.

\subsection{Asymptotic boundary conditions}

For the gravitational field the canonical variables are the spatial metric $g_{ij}$ and their 
conjugate $\pi^{ij}$. The generators
of surface deformation are given by,

\begin{align*}
\mathcal{H}_{\perp} & =\frac{2}{\sqrt{g}}\left(\pi^{ij}\pi_{ij}-\frac{1}{2}\pi^{2}\right)-\frac{1}{2}\sqrt{g}{}^{\left(3\right)}R\approx0\,,\\
\mathcal{H}_{i} & =-2\pi_{i\ |j}^{\ j}\,\approx0.
\end{align*}
Here we have set the cosmological constant equal to zero, and have chosen units such that $8\pi G=1$. 
The deformation parameters that multiply $\mathcal{H}_{\perp}$ and $\mathcal{H}_{i} $ 
in the Hamiltonian are the lapse $N^{\perp}$ and the shift $N^{i}$.

\subsubsection{Power expansion near $\rho=\pm \infty$}
\textit{Schwarzschild as a starting point}

Since our spacelike surfaces are asymptotically null, we must take
as a starting point a coordinate system for the Schwarzschild metric
which incorporates this property. This is provided by the
Eddington-Finkelstein coordinates in terms of which the line element
reads, 
\begin{equation}
ds^{2}=-(dx^{0})^{2}+dr^{2}+r^{2}\left(d\theta^{2}+\sin^{2}\theta d\phi^{2}\right)+\frac{M}{4\pi r}\left(dx^{0} -  dr\right)^{2}\,.\label{eq:metric good coords}
\end{equation}

The next step is to pass to hyperbolic coordinates, through the change
of variables \eqref{emb1}-\eqref{emb2},
extract the asymptotic
form of the resulting expression, and proceed by trial and error as explained in section \ref{qrf}.

In terms of the  the dimensionless radial variable 
\begin{equation}
\rho=\frac{r}{\tau_0},\label{rho}
\end{equation}
the resulting boundary conditions are,

\begin{eqnarray}
g_{\rho\rho} & = & \tau_{0}^{2}\left(\frac{1}{\rho^{2}}+\frac{f_{\rho\rho}^{\left(-4\right)}}{\rho^{4}}+\frac{f_{\rho\rho}^{\left(-5\right)}}{\rho^{5}}+O\left(\rho^{-6}\right)\right)\,,\label{grr}\\
g_{\rho a} & = & \tau_{0}^{2}\left(\frac{f_{\rho a}^{\left(-2\right)}}{\rho^{2}}+\frac{f_{\rho a}^{\left(-3\right)}}{\rho^{3}}+O\left(\rho^{-4}\right)\right)\,,\\
g_{ab} & = & \tau_{0}^{2}\left(\gamma_{ab}\rho^{2}+f_{ab}^{\left(1\right)}\rho+f_{ab}^{\left(0\right)}+\frac{f_{ab}^{\left(-1\right)}}{\rho}+O\left(\rho^{-2}\right)\right)\,,\\
\pi^{\rho\rho} & = & -\sqrt{\gamma}\rho^{3}+p_{\left(1\right)}^{\rho\rho}\rho+p_{\left(0\right)}^{\rho\rho}+O\left(\rho^{-1}\right)\,,\\
\pi^{\rho a} & = & \frac{p_{\left(-1\right)}^{\rho a}}{\rho}+\frac{p_{\left(-2\right)}^{\rho a}}{\rho^{2}}+O\left(\rho^{-3}\right)\,,\\
\pi^{ab} & = & -\sqrt{\gamma}\gamma^{ab}\frac{1}{\rho}+\frac{3\sqrt{\gamma}f^{ab\left(1\right)}}{4\tau_{0}}\frac{1}{\rho^{2}}+\frac{p_{\left(-3\right)}^{ab}}{\rho^{3}}+\frac{p_{\left(-4\right)}^{ab}}{\rho^{4}}+O\left(\rho^{-5}\right)\,.\label{piab}
\end{eqnarray}

The coefficients which are explicitly shown above are those that will
appear in the surface integrals later on. They are not all independent,
but must obey relations among them, in order for the action principle
that will be discussed next be well-defined. They are the demand that
some terms ${\cal H}_{\mu}^{\left(n\right)}$ in the asymptotic expansion
of ${\cal H}_{\mu}$, should vanish strongly, namely: (i) ${\cal H}_{\perp}^{\left(-1\right)}=0$.
This ensures that the symplectic term in the action is finite\footnote{The symplectic term will be taken to be $-\int\dot{\pi}^{ij}g_{ij}\ d^{3}x$.
It is interesting to note that in this momentum representation the
Hamiltonian action is equal to the Hilbert action up to surface terms
at spatial infinity; this phenomenon only happens in four spacetime
dimensions. We have not found a simple way to make finite the conjugate
term $\int\pi^{ij}\dot{g}_{ij}d^{3}x$.}. (ii) ${\cal H}_{a}^{\left(0\right)}={\cal H}_{\rho}^{\left(-2\right)}={\cal H}_{\rho}^{\left(-3\right)}=0$.
Together with (i) conditions (ii) make the surface term
in the variation of the Hamiltonian finite, permit to take the
$\delta$ outside on the left-hand side of \eqref{eq:delta} below,
and give the simple forms \eqref{eq:P}-\eqref{eq:K} for $Q$. These
extra algebraic conditions are harmless. One could have solved them
to express the boundary conditions in terms of a lesser number of
coefficients which would then be all independent, but the resulting
expressions are complicated, and it is more convenient to carry them
along.

\subsubsection{Parity conditions}

In addition to the power law decays \eqref{grr}--\eqref{piab} it is necessary to introduce parity conditions. This is achieved by splitting some of the variables in longitudinal and transverse parts as follows
\begin{equation}
\tau_{0}\tilde{f}_{ab}^{\left(1\right)}=\nabla_{ab}F+\star\nabla_{ab}\bar{G}\,,\label{f1ab}
\end{equation}
\begin{equation}
h_{ab}=\frac{1}{2}\left(\nabla_{ab}N+\star\nabla_{ab}\bar{N}\right) \,\label{habdef},
\end{equation}
Here the news tensor $h_{ab}$, given by 
\begin{equation}
h^{ab}=2\tilde{p}_{\left(-3\right)}^{ab}-\sqrt{\gamma}\tilde{f}^{ab\left(0\right)}\,.\label{eq:hab}
\end{equation}
is the gravitational analog of the electromagnetic $h^a$. 

If one has a symmetric tensor $s_{ab}$,
we denote its trace by $s$, i.e., we use the same letter but without
indices. The traceless part will be denoted with a tilde $\tilde{s}^{ab}=s^{ab}-\frac{1}{2}\gamma^{ab}s$.

In the above equations the operators $\nabla_{ab}$ and $\star\nabla_{ab}$ given by,
\begin{equation}
\nabla_{ab}=2(\nabla_{a}\nabla_{b}+\nabla_{b}\nabla_{a}-\gamma_{ab}\nabla^{2}),\ \ \star\nabla_{ab}=2\sqrt{\gamma}\gamma^{cd}\left(\epsilon_{ac}\nabla_{b}\nabla_{d}+\epsilon_{bc}\nabla_{a}\nabla_{d}\right)\,,\label{nablas}
\end{equation}
are the tensor analogs of the vector gradient, $\nabla_{a}$, and
curl $\star\nabla_{a}$ appearing in \eqref{eq:divcurl}. These operators
were used by Regge and Wheeler in their analysis of the stability
of a Schwarzschild singularity  \cit{Regge:1957td}, and obey the
key properties
\begin{equation}
 \nabla^{ab}\left(\star\nabla_{ab}\right)=\star\nabla_{ab}\left(\nabla^{ab}\right)=0,
 \end{equation}
 when they act on scalar functions, just as their vector counterparts.
Their kernel is spanned by the $\ell=0$ and $\ell=1$ modes of the
corresponding scalar functions on which they act.

The parity conditions will be then the following,
\be{paritygr}
\left. (F, N,\bar N)\right|_{+\infty} =
\left. (F, N,\bar N)\right|_{-\infty} \ \ 
\mbox{for each } (\vartheta,\varphi),
\ee
in close analogy with eq. \eqref{parity} for electromagnetism.

\subsubsection{Preservation of the boundary conditions}

One may verify using Einstein's equations in Hamiltonian form

\begin{align}
\dot{g}_{ij}= & \frac{4}{\sqrt{g}}N^{\perp}\left(\pi_{ij}-\frac{\pi}{2}g_{ij}\right)+N_{i/j}+N_{j/i}\,,\label{eq:EOM1}\\
\dot{\pi}^{ij}= & -\frac{1}{2}N^{\perp}\sqrt{g}\left(^{\left(3\right)}R^{ij}-\frac{1}{2}g^{ij\left(3\right)}R\right)+\frac{N^{\perp}}{\sqrt{g}}g^{ij}\left(\pi^{kl}\pi_{kl}-\frac{1}{2}\pi^{2}\right)-\frac{4N^{\perp}}{\sqrt{g}}\left(\pi_{\;l}^{i}\pi^{jl}-\frac{1}{2}\pi^{ij}\pi\right)\label{eq:EOM2}\\
 & +\frac{1}{2}\sqrt{g}\left(N^{\perp/i/j}-g^{ij}N_{/k/k}^{\perp}\right)+\left(N^{k}\pi^{ij}\right)_{/k}-N_{/k}^{i}\pi^{kj}-N_{/k}^{j}\pi^{ki}\,,\nonumber \\
{\cal H}_{\mu} & =0\,,
\end{align}
that the most general surface deformation that preserves \eqref{grr}-\eqref{piab}
takes the form: 
\begin{eqnarray}
N^{\perp} & = & \epsilon_{\left(1\right)}^{\perp}\rho+\frac{\epsilon_{\left(-1\right)}^{\perp}}{\rho}+O\left(\rho^{-2}\right)\,,\label{eq:lapse}\\
N^{\rho} & = & -\frac{\epsilon_{\left(1\right)}^{\perp}}{\tau_{0}}\rho^{2}-\frac{1}{2}\left(\nabla_{a}\epsilon_{\left(0\right)}^{a}\right)\rho+\epsilon_{(0)}^{\rho}+O\left(\rho^{-1}\right)\,,\\
N^{a} & = & \epsilon_{\left(0\right)}^{a}-\frac{1}{\tau_{0}}\left(\nabla^{a}\epsilon_{\left(1\right)}^{\perp}\right)\frac{1}{\rho}+\frac{\epsilon_{(-2)}^{a}}{\rho^{2}}+O\left(\rho^{-3}\right)\,,\label{eq:shift}
\end{eqnarray}
where,

\begin{eqnarray*}
\epsilon_{\left(-1\right)}^{\perp} & = & \epsilon_{\left(1\right)}^{\perp}\left[\frac{1}{2}\frac{p_{\left(-3\right)}}{\sqrt{\gamma}}-\frac{1}{4}f^{\left(0\right)}+\frac{1}{4}f^{ab\left(1\right)}f_{ab}^{\left(1\right)}\right]+\frac{1}{4}\nabla^{2}\epsilon_{\left(1\right)}^{\perp}\,,\\
\epsilon_{\left(0\right)}^{\rho} & = & \frac{\epsilon_{\left(1\right)}^{\perp}}{\tau_{0}}\left[\frac{1}{4}\left(f^{\left(0\right)}-f^{ab\left(1\right)}f_{ab}^{\left(1\right)}-2f_{\rho\rho}^{\left(-4\right)}\right)+\frac{1}{\sqrt{\gamma}}\left(p_{\left(1\right)}^{\rho\rho}-\frac{1}{2}p_{\left(-3\right)}\right)\right]+\frac{1}{4\tau_{0}}\nabla^{2}\epsilon_{\left(1\right)}^{\perp}\,,\\
\epsilon_{\left(-2\right)}^{a} & = & -\frac{1}{4}\partial^{a}\left(\nabla_{b}\epsilon_{\left(0\right)}^{b}\right)-\frac{\epsilon_{\left(1\right)}^{\perp}}{\tau_{0}}\gamma^{ab}f_{\rho b}^{\left(-2\right)}+\epsilon_{\left(1\right)}^{\perp}\frac{2}{\tau_{0}}\frac{p_{\left(-1\right)}^{\rho a}}{\sqrt{\gamma}}+\frac{1}{2\tau_{0}}f^{ab\left(1\right)}\partial_{b}\epsilon_{\left(1\right)}^{\perp}\,.
\end{eqnarray*}

The coefficients in the asymptotic expansion above are written explicitly
up to the order in which they appear in the surface integrals. One
sees that they are determined by the functions $\epsilon_{\left(1\right)}^{\perp}(\vartheta,\varphi)$
and $\epsilon_{\left(0\right)}^{a}(\vartheta,\varphi)$ which will correspond
to supertranslations and Lorentz transformations respectively. The
latter appear in $\epsilon_{\left(0\right)}^{a}$ through the decomposition

\begin{eqnarray}
\epsilon_{\left(0\right)}^{a} & = & \partial^{a}\Lambda+\frac{\epsilon^{ab}}{\sqrt{\gamma}}\partial_{b}\Omega\ ,\label{eq:epsiloncero}
\end{eqnarray}
with
\begin{equation}
\Lambda=\vec{\beta}\cdot\hat{r},\ \ \Omega=\vec{\omega}\cdot\hat{r},\label{eq:lambda}
\end{equation}
where $\vec{\beta}$ is the parameter of an infinitesimal boost, and
$\vec{\omega}$ is the vector angle of an infinitesimal spatial rotation\footnote{``Superrotations'' have been introduced in  \cit{Banks:2003vp,Barnich:2009se,Barnich:2010eb}.
They correspond to $\epsilon_{\left(0\right)}^{a}$ singular on the
sphere.}.

In order for the parity conditions \eqref{parity} to be preserved in time one must demand 
\begin{equation}
 \epsilon_{\left(1\right)}^{\perp} (+\infty)=\epsilon_{\left(1\right)}^{\perp} 
(-\infty).
\end{equation}
In addition, just as in electromagnetism an infinite sequence of additional 
conditions appears. They present no problem here either.

\subsubsection{Surface deformation algebra}
\label{sdaa}

Specializing the general surface deformation algebra \eqref{def2}
to two deformations $\xi_{1}=\xi$ and $\xi_{2}=\eta$ of the form
\eqref{def1}, \eqref{def2} one finds for the commutator $\xi_{12}=\zeta$
\begin{eqnarray}
\zeta_{\left(1\right)}^{\perp} & =&\xi_{\left(0\right)}^{a}\partial_{a}\eta_{\left(1\right)}^{\perp}-\eta_{\left(0\right)}^{a}\partial_{a}\xi_{\left(1\right)}^{\perp}+\frac{1}{2}\left(\xi_{\left(1\right)}^{\perp}\nabla_{a}\eta_{\left(0\right)}^{a}-\eta_{\left(1\right)}^{\perp}\nabla_{a}\xi_{\left(0\right)}^{a}\right), \label{sda1}\\
\zeta_{\left(0\right)}^{a} & =&\xi_{\left(0\right)}^{b}\partial_{b}\eta_{\left(0\right)}^{a}-\eta_{\left(0\right)}^{b}\partial_{b}\xi_{\left(0\right)}^{a}\,, \label{sda2}
\end{eqnarray}
and that $\zeta^{a}$ closes in terms of $\xi^{a}$ and $\eta^{a}$
according to the Lorentz algebra.

These equations are the BMS algebra in the standard form (see for
example  \cit{Sachs:1962zza}). 

\subsubsection{Supertranslation memory}

Consider a time translation:
\begin{equation}
  \epsilon_{\left(1\right)}^{\perp}=1.
\end{equation} 
The equation of motion \eqref{eq:EOM1} yields then 

\begin{equation}
  \dot{\tilde{f}}_{ab}^{(1)}=2h_{ab}, \label{dotfab}
\end{equation}
which implies 
\begin{equation}
  \dot{F}=N. \label{cien}
\end{equation}
Therefore, when time $\delta t$ elapses a supertranslation of magnitude
\begin{equation}
  \delta F=N \delta t,
\end{equation}
takes place. This is the supertranslation memory effect, analogous to the fiber 
memory of electromagnetism discussed in section \ref{sec:Poinc}.

\subsection{Electric and magnetic BMS charges}

We saw in the electromagnetic case that it was necessary to employ, asymptotically on the hourglass an 
electric-magnetic duality invariant formalism, in order to be able to improve the
generators. The same 
will occur in gravitation. In that case we do not possess at the moment an explicit 
electric-magnetic duality invariant description of the linearized theory on the hourglass, which is what is needed at large distances. However, it is reasonable to 
assume that such a description exists, and that it can be constructed along lines similar to those employed succesfully for asymptotic 
planes  in [\citenum{Henneaux:2004jw, Bunster:2006rt}].

Fortunately, it turns out that assuming the existence of the asymptotic electric-magnetic duality invariant description, one 
can conjecture by analogy some of the elements that are needed. The coherence of the results thus obtained reinforces the hypothesized  
existence of the electric-magnetic representation. We now pass to discuss those elements.

\subsubsection{Electric BMS charge}

If one varies the Hamiltonian  
\begin{equation}
H_{0}=\int d^{3}x\left(N^{\perp}\mathcal{H}_{\perp}+N^{i}\mathcal{H}_{i}\right)\,,
\end{equation}
in the electric representation, with the Lorentz parameters $\vec \omega$, $\vec\beta$ set equal to zero,
one finds
\begin{equation}
\delta H_{0}=-\oint d\vartheta d\varphi\epsilon_{\left(1\right)}^{\perp}\left(\vartheta,\varphi\right)\left.\left[\delta\mathcal{P}+\frac{1}{2}h^{ab}\delta\left(\tau_{0}f_{ab}^{\left(1\right)}\right)\right]\right|_{-\infty}^{+\infty},
\label{eq:delta}
\end{equation}
with
  \begin{equation}
{\cal P}\left(\vartheta,\varphi\right)=\tau_{0}\sqrt{\gamma}\left(\nabla^{a}f_{\rho a}^{\left(-2\right)}
+3f_{\rho\rho}^{\left(-5\right)}+\frac{3}{2}f^{\left(-1\right)}
+\frac{1}{\sqrt{\gamma}}\left(\frac{1}{2}h^{ab}f_{ab}^{\left(1\right)}
+2p_{\left(-4\right)}\right)\right)\,.\label{eq:P}
\end{equation}
This identifies $Q\left(\vartheta,\varphi\right)$
 \be{Qgrav}
 Q=\left. \mathcal{P}\right|^{+\infty}_{-\infty},
 \ee
 as the (electric) supertranslation charge\footnote{
We have grouped $\tau_{0}$ together with $f_{ab}^{\left(1\right)}$
because the product $\tau_{0}f_{ab}^{\left(1\right)}$ is the analog
of the electromagnetic $a_{a}^{\left(0\right)}$. They both have the
property of remaining finite in the limit $\tau_{0}\rightarrow0$
which turns the hyperboloids into light cones. Conversely, the news
$h^{ab}$, whose formula \eqref{eq:hab} does not have an explicit
$\tau_{0}$ in it is the analog of the electromagnetic news $h^{a}$
whose definition \eqref{ha} does have a $\tau_{0}$ in it. It also
remains finite in the limit $\tau_{0}\rightarrow0$. These differences
in the units between the electromagnetic and the gravitational case
are produced by the choice $8\pi G=1$.  Again in the above formulas, one should include $\tau_{0}$ into each
coefficient to have the natural variables.}$^{,}$\footnote{It is interesting to note the presence of the quadratic term $h^{ab}f_{ab}^{\left(1\right)}$
in ${\cal P}$. In a situation in which one has gravitational radiation
being emitted by a confined source it would represent interference
between the radiation field and the field that remains bound
to the source (``near field''). This interference contribution is
not emitted, but it remains bound to the source because it is part
of ${\cal P}$. A similar phenomenon happens in electromagnetism and
it occurs there in the volume integral of ${\cal H}_{\mu}^{\left(elm\right)}$.
This has been discussed in  \cit{Teitelboim:1970mw}.}. In the analogy with 
 electromagnetism, the $l=0$ and $l=1$ of the charge \eqref{Qgrav} correspond
  to spacetime 
 translations whereas those with $l\geq 2$ correspond to the electromagnetic 
 charges with spherical modes $l\geq 1$.
The first term on the right hand side of \eqref{eq:delta} may be compensated in the 
standard manner by defining a partially improved Hamiltonian $\tilde{H}_{0}^{\text{\tiny{elec}}}$ through
\be{H0tilde}
\tilde{H}_{0}^{\text{\tiny{elec}}}=H_{0}+\oint\epsilon_{\left(1\right)}^{\perp}\left(\vartheta,\varphi\right)Q\left(\vartheta,\varphi\right).
\ee
The Hamiltonian \eqref{H0tilde} is the analog of the Maxwell electric Hamiltonian for spacetime 
translations {\it and} improper gauge transformations, and just as that one it will 
need to be improved to eliminate the surface term 
\be{niexp}
\oint \frac{1}{2}\epsilon_{\left(1\right)}^{\perp}h^{ab}\delta\left(\tau_{0}.\tilde f_{ab}^{\left(1\right)}\right)\bigg|_{-\infty}^{+\infty}=\oint\frac{1}{4}\gamma^{\frac{1}{2}}\epsilon_{\left(1\right)}^{\perp}\left(\nabla_{ab}N+\star\nabla_{ab}\bar{N}\right)\left(\nabla_{ab}\delta 
F+\star\nabla_{ab}\delta\bar{G}\right)\bigg|_{-\infty}^{+\infty}.
\ee
The term proportional to $\delta F$ on the right hand side of eq. \eqref{niexp} vanishes 
when the parity conditions hold but the one proportional to $\delta \bar{G}$, which reads
\be{dggr}
\delta H_0=-\oint \bar\eta \delta \bar G,
\ee
with
\be{eta}
\bar\eta = \frac{1}{4}\nabla^{ab}\left[\epsilon^\perp_{(1)}\left(\nabla_{ab}\bar N - \star\nabla_{ab}N\right)\right]
\ee
does not.

\subsubsection{Magnetic BMS charges}

In order to eliminate \eqref{dggr} one should supplement the Hamiltonian acting with the generator of magnetic BMS transformations, whose form we do not know, but which should be such that the surface term in its variation should read
\be{stv}
-\oint\bar\epsilon^\perp_{(1)}\delta\bar{\cal P}  ,
\ee
where, by definition, $\bar{\cal P}$ is the magnetic BMS charge and $\bar\epsilon^\perp_{(1)}$ is the magnetic deformation parameter, so we must have,
$$
\oint\bar\epsilon^\perp_{(1)}\delta\bar{\cal P} = \oint \bar\eta \delta \bar G,
$$
and the question is: what is the relationship between $\bar{\cal P}$ and $\bar G$?

This can be established by recalling from electromagnetism that one would like the parameter $\eta$ to bring the magnetic memory. So we set
\be{epsone}
\epsilon^\perp_{(1)}=1,
\ee
in which case the boundary term reads
\be{bt2}
\delta H_0= -\oint \nabla^4\bar N\delta (\gamma^{\frac{1}{2}}\bar G)\bigg|_{-\infty}^{+\infty}
= -\oint \bar N\delta \left( \nabla^4(\gamma^{\frac{1}{2}}\bar G)\right)\bigg|_{-\infty}^{+\infty},
\ee
where
\be{nabla4}
\nabla^{4}=\nabla_{ab}\nabla^{ab}=\star\nabla_{ab}\star\nabla^{ab}=8\left[\left(\nabla^{2}\right)^{2}+2\nabla^{2}\right].
\ee
Comparison with the magnetic analog of Eq. \eqref{cien} then gives
\be{pbar}
\bar{\cal P}= \gamma^{\frac{1}{2}}\nabla^4 \bar G
\ee
The identification \eqref{pbar} will have a significative consistency check when we discuss angular momentum below.

In order to account for the $l=0$ and $l=1$ modes of $\bar{\cal P}$ (electric and magnetic
translations generators) one needs to bring in Dirac strings into $\bar G$ because those 
modes are in the kernel of $\nabla^4$.

\subsection{Lorentz generators}

If one works solely in the electric representation one 
finds that if one considers the motion corresponding to a Lorentz transformation, 
with infinitesimal rotation and boost parameters $\vec{\omega}$ and 
$\vec{\beta}$, one must improve the Hamiltonian by adding to it the surface term, 
\[
\vec{\omega}\cdot\vec{J}_{\mbox{\tiny el}}+\vec{\beta}\cdot\vec{K}_{\mbox{\tiny el}},
\]
with,
\begin{eqnarray}
\vec{J}_{\mbox{\tiny el}} & = & \oint 2\tau_{0}^{2}\epsilon^{ab}\left(\partial_{b}\hat{r}\right)\left(\frac{p_{\:\,a\left(-2\right)}^{\rho}}{\sqrt{\gamma}}-f_{\rho a}^{\left(-3\right)}\right)\,,\label{eq:J}\\
\vec{K}_{\mbox{\tiny el}} & = & \oint \tau_{0}^{2}\sqrt{\gamma}\left\{ \hat{r}\left(1+f^{\left(0\right)}+2f_{\rho\rho}^{\left(-4\right)}+\frac{p_{\left(-3\right)}}{\sqrt{\gamma}}\right)+2\left(\partial^{a}\hat{r}\right)\left(\frac{p_{\:\,a\left(-2\right)}^{\rho}}{\sqrt{\gamma}}-f_{\rho a}^{\left(-3\right)}\right)\right\} .\label{eq:K}
\end{eqnarray}
When this is done the generators are well-defined. No additional surface 
integral containing the news, analogous to \eqref{niexp} appears. This is quite reasonable because the Lorentz motion lies within the hourglass. 
The first term $\tau_{0}^{2}\sqrt{\gamma}\hat{r}$
in \eqref{eq:K}, which has no variation, has been incorporated so
that for Minkowski space the numerical value of the boost generator
is zero. 

The electric generators thus obtained are the analog of the electromagnetic 
angular momentum \eqref{electricJ}. 

If one takes $\epsilon^\perp_{(1)}=1$, and evaluates the rate of change of $\vec J_{el}$ , one finds, either by direct calculation from Einstein's equations or, better, by using Eq. \eqref{qdotgrav} below,
\be{djeldt}
\frac{d\vec J_{el}}{dt}=-\oint (\gamma^{\frac{1}{2}}\nabla^4 \bar G)\bar N_{,a} \vec\xi^a\bigg|_{-\infty}^{+\infty} .
\ee
Eq. \eqref{djeldt} is the analog of \eqref{dtJ} for electromagnetism. It provides a consistency check of the definition \eqref{pbar} because if we bring in the magnetic analog  $\bar F$ of $F$, and postulate the magnetic memory equation,
\be{maganalog}
\dot {\bar F}= \bar N,
\ee
then,
\be{Jcons}
\vec J= \vec J_{el}+ \oint \bar {\cal P}\bar F_{,a} \vec\xi^a\bigg|_{-\infty}^{+\infty}
\ee
is conserved
\be{djdtgr}
\frac{d\vec J}{dt}=0,
\ee
when the parity conditions hold.

So, by appealing to electric-magnetic duality one can find a conserved angular momentum in general relativity, even in the presence of radiation, but provided the net radiation flux is zero. 

For boosts one must include an extra term (see comment at the end of the next subsection). Thus one has in general,
\be{hbl}
H_{\mbox{\tiny Lorentz}}= H_{\mbox{\tiny Lorentz}}^{\mbox{\tiny el}}+ \oint  \bar Q\left(\xi^a\partial_aF-\frac{3}{2}\nabla_a\xi^aF \right) .
\ee 
(The second term $\nabla_a\xi^a$ vanishes for rotations).

\subsection{Symmetry algebra}

The analog of \eqref{QHcom} and \eqref{eq:QP} for electromagnetism
is 
\begin{eqnarray}
\left[Q^M\left(\vartheta,\varphi\right),Q^N\left(\vartheta',\varphi'\right)\right]^{\star} &=&0\nonumber \\
\left[Q^M\left(\vartheta,\varphi\right),\vec{J}\right]^{\star} &=&\partial_{a}\left(\vec{\xi}_{\text{R}}^{a}Q^M\left(\vartheta,\varphi\right)\right) \label{BMSalgebra}\nonumber\\
\left[Q^M\left(\vartheta,\varphi\right),\vec{K}\right]^{\star} &=&\partial_{a}\left(\vec{\xi}_{B}^{a}Q^M\left(\vartheta,\varphi\right)\right) \\ &+&\frac{3}{2}Q^M\left(\vartheta,\varphi\right)\left(\nabla_{a}\vec{\xi}_{B}^{a}\right),\nonumber
\end{eqnarray}
while the Lorentz generator $\vec{K}$ and $\vec{J}$ close among themselves in 
the Lorentz algebra.

We have used Dirac brackets $\left[\;,\;\right]^{\star}$ here because,
as explained in the introduction it is only through them that the
surface term alone can act as a generator. If one wanted to use Poisson
brackets one would have to add to the surface term the weakly vanishing
volume part of the generator.

For the electric generators $Q$ and the Lorentz generators, equations \eqref{BMSalgebra} can be obtained directly from the  surface deformations algebra \eqref{sda1} and \eqref{sda2}. It is then extended to the magnetic generators by duality.

\subsection{Emission and absorption rates. Charge memory}

\subsubsection{General formula for emission rates}

In this case we only possess the formula stemming from electric sector, that is,
\be{qdotgrav}
\dot Q_\alpha= \oint \frac{1}{2}h^{ab}\delta_\alpha\left(\tau_{0}\tilde f_{ab}^{\left(1\right)}\right) \Big|_{-\infty}^{+\infty}.
\ee
The analog of the second expression on the right hand side of \eqref{rates} is not obvious to guess because, this time, under a duality transformation, one must turn the electric time  into magnetic time. That is, one would have to compare motions that have $\epsilon^\perp_{(1)}=1$, $\bar  \epsilon^\perp_{(1)}=0$ with those with $\epsilon^\perp_{(1)}=0$, $\bar  \epsilon^\perp_{(1)}=1$ .

By applying \eqref{qdotgrav} one obtains the following results.

\subsubsection{Electric BMS charges}

\begin{align}
\frac{\partial Q}{\partial t}&=\left(-\frac{1}{\sqrt{\gamma}}h^{ab}h_{ab}+\nabla_{a}\nabla_{b}h^{ab}\right)\bigg|^{+\infty}_{-\infty} \,,\nonumber \\
&= -\frac{1}{4}\gamma^{\frac{1}{2}}\bigg((\nabla^{ab}N)(\nabla_{ab}N) + (\nabla^{ab}\bar N)(\nabla_{ab}\bar N) + \nonumber \\ &+ 2(\nabla^{ab}N)(\star\nabla_{ab}\bar N) +
+\gamma^{\frac{1}{2}}\nabla^4 N\bigg)\bigg|^{+\infty}_{-\infty}, \label{Aeq}
\end{align}
(This equation,  as well as its relationship with the emission rate, were found previously in \cit{Barnich:2009se,Barnich:2011mi}; but they interpreted it as meaning that the Hamiltonian cannot be improved if
   $h_{ab}\neq 0$.)

\subsubsection{Magnetic BMS charge}

The (electric) time derivative of the magnetic BMS charge cannot be obtained from the purely electric sector formula \eqref{qdotgrav}, although $\bar {\cal P}$ does appear in the electrir sector. One must resort to its definition \eqref{pbar} and to the equation of motion \eqref{dotfab}. This yields,
\be{Beq}
\frac{\partial\bar Q}{\partial t}=\gamma^{\frac{1}{2}}\nabla^4
 \bar N\Big|^{+\infty}_{-\infty}
\ee

Note that there is no symmetry between the rates of change of $Q$ and $\bar Q$. That is quite alright because one should not expect any: the duality counterpart of \eqref{Aeq} should be the rate of change of $\bar Q$ with respect to a {\it magnetic} time displacement with $\bar\epsilon^\perp_{(1)}=1$, and  $\epsilon^\perp_{(1)}=0$.

In the same vein, one could define a variable $G$ thorough
\be{calP}
\gamma^{\frac{1}{2}}\nabla^4  G = \cal P.
\ee
Then the derivative of $G$ with respect to electric time would not be equal to $N$ as one can see from \eqref{Aeq}. However, one would expect its derivative with respect to magnetic time to be given by $\bar N$, in analogy with \eqref{Beq}.

\subsubsection{Angular momentum}

One may write, in analogy with \eqref{dtJQ} in electromagnetism

\begin{equation}
\frac{d\vec{J}}{dt}=\oint \dot {\cal P}_M F^M_{\ ,a}\vec \xi^a \label{dJdtgrav}
\end{equation}
where $\dot {\cal P}$ and $\dot{\bar {\cal P}}$ are given by \eqref{Aeq} and \eqref{Beq}.

It should be stressed that formula \eqref{dJdtgrav} has not been proven, but just conjectured by analogy and ``informed guess".
Only the conservation of  $\vec J$ when $\dot {\cal P}$ and $\dot{\bar {\cal P}}$ vanish has been proven (once \eqref{maganalog} has been postulated!). This is because, in the lack of a complete asymptotic two potential theory, we do not posses an analog of the second expression on the right hand side of \eqref{rates}.  But the presumption is  that \eqref{dJdtgrav} will survive the complete development of the asymptotically duality invariant description.

All the comments made for the electromagnetic case in connection with the angular momentum and with its rate of change apply here as well.
   
\subsection{Special solutions}

There are two fundamental solutions of Einstein's equations for which is important to verify that they fit into the present treatment. They are Taub-Nut space and the Kerr solution. We pass to discuss them now.
\subsubsection{Taub--Nut space as a magnetic pole}

The gravitational analog of a magnetic pole in electromagnetism is Taub-Nut space. We will now show that it satisfies our boundary 
conditions, and its magnetic pole nature will be distinctly brought out.

The Taub-NUT metric in Schwarzschild coordinates,
\begin{equation}
ds^{2}=-V\left(r\right)\left[dt+2N\left(1-\cos\vartheta\right)d\varphi\right]^{2}+V\left(r\right)^{-1}dr^{2}+\left(r^{2}+N^{2}\right)\left(d\vartheta^{2}+\sin^{2}\vartheta d\varphi^{2}\right)\,,\label{TN}
\end{equation}
with
\[
V\left(r\right)=1-\frac{2\left(N^{2}+\frac{M}{8\pi}r\right)}{r^{2}+N^{2}}\,,
\]
may be brought by a change of coordinates to obey our boundary conditions with 
\begin{equation}
F=0\,,\label{eq:F-2}
\end{equation}
 and 
\begin{equation}
\bar{G}=-N\log\left(1+\cos\vartheta\right)\, .\label{G}
\end{equation}
These are exactly the expressions \eqref{Gpole} of electromagnetism
for a magnetic pole of charge $g=-N$, with the Dirac string going
through the south pole\footnote{When one discusses Taub-NUT on surfaces which are asymptotically planes,
as it was done in  \cit{Bunster:2006rt}, one finds, that in order
to satisfy the Regge-Teitelboim boundary conditions which include
a parity requirement, one must take half of the string to come out
of the south pole and the other half to come out from the north pole.
No such requirement is present here, where one can take just one string
going out through any point on the sphere. }.

To arrive at \eqref{eq:F-2} and \eqref{G} one takes the following
steps: 

(i) Pass to the analog of the Eddington-Finkelstein coordinates
\begin{align*}
t & =x^{0}+r^{*}-r+\frac{2N^{2}\cot^2\theta}{r}\quad,\quad\vartheta=\theta+\frac{2N^{2}\cot\theta\csc^{2}\theta}{r^{2}}\quad,\quad\varphi=\phi+\frac{2N\cot\theta\csc\theta}{r}\,,
\end{align*}
where $r^{*}$ is the ``tortoise'' radial coordinate, 
\begin{equation}
r^{*}=\int^{r}\frac{dr}{V\left(r\right)}.\label{tortoise}
\end{equation}

(ii) Pass to hyperbolic coordinates according to \eqref{emb1} and \eqref{emb2}.
The resulting expression is complicated in closed form but we only
need the fact that its asymptotic form fits the boundary conditions
\eqref{grr}-\eqref{piab}. One finds that this is indeed the case,
and, in particular, 
\begin{equation}
\tau_{0}f_{\theta\phi}^{\left(1\right)}=-2N\left[1-2\cos\theta+\cos^{2}\theta\right]\,.\label{eq:ftaubnut}
\end{equation}

(iii) Identify from \eqref{eq:ftaubnut} $F$ and $G$ through the
decomposition \eqref{f1ab}, obtaining \eqref{eq:F-2} and \eqref{G}.

\subsubsection{Kerr solution}

  To show that the Kerr solution satisfies our boundary conditions \eqref{grr}-\eqref{piab}, one performs the following steps:

(i) Write the solution in Kerr-Schild coordinates:
\begin{align*}
ds^{2}=&-(dx^{0})^{2}+dx^{2}+dy^{2}+dz^{2}\\
&+\frac{MR^{3}}{4\pi\left(R^{4}+a^{2}z^{2}\right)}\left[-dx^{0}+\frac{R\left(xdx+ydy\right)-a\left(xdy-ydx\right)}{R^{2}+a^{2}}+\frac{zdz}{R}\right]^{2}\,,
\end{align*}
where $R$ is given by 
\[
R=\frac{1}{\sqrt{2}}\left(\sqrt{x^{2}+y^{2}+z^{2}-a^{2}+\sqrt{4a^{2}z^{2}+\left(x^{2}+y^{2}+z^{2}-a^{2}\right)^2}}\right)\,.
\]

(ii) Perform the standard change of basis from Cartesian to spherical
coordinates $r,\theta,\phi$, and then pass to hyperbolic coordinates
by using the change of coordinates \eqref{emb1}, \eqref{emb2} to obtain
the asymptotic form \eqref{grr}-\eqref{piab}. 

One can then evaluate the charges. One find that the only non-vanishing ones are 
\begin{align*}
P^{0} & =\oint {\cal P}=M\,,\\
J_{z} & =aM\,.
\end{align*}
Here, the value of the angular momentum has been calculated using the electric flux integral \eqref{eq:J}. This is alright because the magnetic contribution vanishes for it.

\section*{Acknowledgments}

We express our gratitude to Professor Anna Ceresole for her continued encouragement throughout this work.
The Centro de Estudios Cient\'ificos (CECs) is funded by the Chilean
Government through the Centers of Excellence Base Financing Program
of Conicyt. C.B. wishes to thank the Alexander von Humboldt Foundation
for a Humboldt Research Award. The work of A.P. is partially funded
by Fondecyt Grants N\textsuperscript{o} 1171162 and N\textsuperscript{o} 1181496. 

\section*{{\large{}Appendices}}

\global\long\def\thesection{\Alph{section}}
 \setcounter{section}{0}


\section{Poincar\'e generators}\label{app1}

In this appendix we give the expressions for the Killing vectors of
the Poincar\'e group in the foliation by hyperboloids with varying center
and fixed radius. The transformations from Minkowskian coordinates $(x^{0},x^{1},x^{2},x^{3})$ to the coordinates  $(t,r,\vartheta,\varphi)$ adapted to the hyperbolic hourglass slicing is given in \eqref{emb1}, \eqref{emb2}. The metric in these coordinates reads
\begin{eqnarray}
ds^2 &=& -\left(1+\frac{r^2}{\tau_0^2}\right)dt^2 +  r^2 d\Omega^2 + \nonumber \\
 && +\frac{1}{\left(1+\frac{r^2}{\tau_0^2}\right)}
\left(dr- \sqrt{\frac{r^2}{\tau_0^2}\left(1+\frac{r^2}{\tau_0^2}\right)}\ dt\right)^2. \label{gretadm}
\end{eqnarray}

We use the following notation: The components of a four-vector
$v^{\mu}$ referred to the Minkowski coordinate system are grouped as $v^{\mu}=(v^{0},\vec{v})$, while the boldface refers
to vector fields defined in terms of their components with respect
to the hyperbolic hourglass foliation. In this notation
the rotation generators around the three spatial axes, that leave
the origin $x^{i}=0$ invariant for any fixed $t$, 
\begin{eqnarray}
\boldsymbol{\xi}_{R}^{1}={\bf J}_{23} & = & -\sin\varphi\mbf{\pa}_{\vartheta}-\cot\vartheta\cos\varphi\mbf{\pa}_{\varphi}\,,\label{r1}\\
\boldsymbol{\xi}_{R}^{2}={\bf J}_{31} & = & \cos\varphi\mbf{\pa}_{\vartheta}-\cot\vartheta\sin\varphi\mbf{\pa}_{\varphi}\,,\label{phirot}\\
\boldsymbol{\xi}_{R}^{3}={\bf J}_{12} & = & \mbf{\pa}_{\varphi}\,,\label{r3}
\end{eqnarray}
are written as 
\begin{equation}
\vec{\boldsymbol{\xi}}_{R}=\hat{r}\times\vec{\boldsymbol{\nabla}}=\gamma^{-\frac{1}{2}}\epsilon^{ab}\nabla_{b}\hat{r}\ \mbf{\pa}_{a}\,.\label{compact}
\end{equation}

The boosts generators $\boldsymbol{\xi}_{B}^{i}={\bf J}_{i0}$ along
the three spatial directions which leave the center of the hyperboloid
$t=$constant fixed are given by 
\begin{equation}
\vec{\boldsymbol{\xi}}_{B}=\sqrt{1+\frac{\tau_0^2}{r^2}} \left(\frac{}{}r\hat{r} \mbf{\partial}_{r}+ \nabla^{a}\hat{r}\ \mbf{\pa}_{a}\right) \,.\label{boost}
\end{equation}
These spatial rotations and boosts map a given hyperboloid $t=$constant
onto itself. For $r\rightarrow\infty$ the tangential part of the boost generators \eqref{boost} together with the rotations \eqref{compact}
close among themselves and form a realization of the Lorentz group on the two-sphere.

The translations do not map the hyperboloid onto itself. Their generators
are 
\begin{equation}
\mbf{\xi}_{T\mu}={\bf P}_{\mu}=\left(\mbf{\partial}_{t},\vec{\boldsymbol{\nabla}}\right)\,,\label{trans}
\end{equation}
or, expressed in terms of normal and tangential components, 
\begin{eqnarray}
\mbf{\xi}_{T^0} & = & \sqrt{1+\frac{r^2}{\tau_0^2}}\left( {\bf n}-\left|\frac{r}{\tau_0}\right|\mbf{\partial}_{r}\right) , \label{tt} \\
\mbf{\vec{\xi}}_{T} & = & -\hat{r}\left|\frac{r}{\tau_0}\right|{\bf n}+\left(1+\frac{r^2}{\tau_0^2}\right)\hat{r}\mbf{\partial}_{r}+\frac{1}{r}(\nabla^{a}\hat{r})\ \mbf{\pa}_{a}\ .
\end{eqnarray}

Here, \textbf{n} is the future oriented unit normal to the hyperboloids $t$=constant. The (future directed) unitary vector ortogonal to them is,
which is given by
\begin{equation}
{\bf n}=\left(1+\frac{r^2}{\tau_0^2}\right)^{-\frac{1}{2}}\mbf{\partial}_{t}+\left|\frac{r}{\tau_0}\right|\mbf{\partial}_{r}\,.\label{n}
\end{equation}


\section{Initial data for the hyperbolic hourglass}

\label{app2}

The parity conditions for $N$ and $\bar N$ say that the value of the electromagnetic news vector $h^a$ at $r\rightarrow\infty$ is equal to  its value at $r\rightarrow-\infty$,
\be{cc1}
\Psi^a_{(2)}=h^a(+\infty) - h^a(-\infty)=0,
\ee
where the subindex $(2)$ is used here to indicate that the condition contains coefficients of order
 $\O(r^{-2})$ in the expansion of the fields. Once this condition is imposed, one must demand that it be
  preserved under Poincar\'e transformations
and gauge transformations, proper and improper. Being a vector defined on the sphere at infinity
 out of gauge invariant quantities, it is evident that $\Psi^a_1=0$ is invariant under Lorentz and gauge transformations. One only needs to be concerned with spacetime translations. In view of the Lorentz invariance, it is sufficient to consider only time translations, and demand that
\be{cc2}
\Psi^a_{(3)}\equiv\dot \Psi^a_{(2)} =0.
\ee
In what follows  we will set, for simplicity, $\tau_0=1$. Using the equations of motion \eqref{dA-1}, \eqref{dpi-1} with $(\xi^\perp, \xi^i)$ being the components of the time translation Killing vector \eqref{tt}, one finds that the expansion coefficients of the different fields satisfy,
\begin{eqnarray}
\dot a_{a}^{(n)} &=& \gamma^{-\frac{1}{2}}\gamma_{ab}\left( \pi^b_{(n)} + \pi^b_{(n+2)}\right)  + \hspace{-0.3cm} \sum_{0\leq 2m\leq n} 
\hspace{-0.3cm} b_m f_{ar}^{(n-2m+2)}, \\
\dot a_{r}^{(n)} &=& \gamma^{-\frac{1}{2}}\pi^r_{(n-2)}, \\
\dot \pi^{r}_{(n)} &=& \partial_a\left[\gamma^{\frac{1}{2}}\gamma^{ab}\left( f_{br}^{(n)} + f_{br}^{(n+2)}\right)  + \hspace{-0.3cm} \sum_{0\leq 2m\leq n} 
\hspace{-0.3cm} b_m \pi^{a}_{(n-2m+2)}\right], \\
\dot \pi^{a}_{(n)} &=& (n-1)\left[\gamma^{\frac{1}{2}}\gamma^{ab}\left( f_{br}^{(n-1)} + f_{br}^{(n+1)}\right)  + \hspace{-0.5cm} \sum_{0\leq 2m\leq n-1} 
\hspace{-0.3cm} b_m \pi^{a}_{(n-2m+1)}\right] \nonumber\\ 
&+& 
\partial_b\left(\gamma^{\frac{1}{2}}\gamma^{cb}\gamma^{da}f_{cd}^{(n-2)}\right),
\end{eqnarray}
where $b_n$ is the $n$th coefficient in the expansion of $\sqrt{1+x}$. 

For  $F_{ar}$, which will be of importance below, one has
\begin{eqnarray}
	\dot f_{ar}^{(n)} &=& (n-1)\left[\gamma^{-\frac{1}{2}}\gamma_{ab}\left(\pi^b_{(n-1)} +\pi^b_{(n+1)}\right) + \hspace{-0.3cm} \sum_{0\leq 2m\leq n-1} \hspace{-0.3cm} b_m  f_{ar}^{(n-2m+1)}\right]
	\nonumber \\ 
&+& \ \ \	\partial_a\left(\gamma^{-\frac{1}{2}}\pi^r_{(n-2)}\right).
\end{eqnarray}

From this expressions one finds that  \eqref{cc2} reads
\be{psi3}
\Psi_{(3)}^a= \left.\left[2\left(\pi^a_{(3)} +\gamma^{\frac{1}{2}}\gamma^{ab}f_{br}^{(3)}\right) + \gamma^{\frac{1}{2}}\gamma^{ab}\partial_b\left(\gamma^{-\frac{1}{2}}\pi^r_{(0)}\right) + \partial_b\left(\gamma^{\frac{1}{2}}\gamma^{cb}\gamma^{da}f_{cd}^{(0)}\right)\right]\right|^{+\infty}_{-\infty}=0
\ee
Note that the highest order coefficients entering $\Psi_{(3)}^a$ are the ones of  $\O(r^3)$ in the transverse fields.
This  shows that one is free to give $\pi^r_{(0)}$ and $f_{\vartheta\varphi}^{(0)}$ , together with
 $\pi^a_{(3)}$ and $f_{ar}^{(3)}$ on one half of the hourglass, say $r\rightarrow +\infty$,  and Eq. \eqref{psi3} will relate them with the corresponding ones on its incoming image, $r\rightarrow -\infty$.

This phenomenon continues indefinitely, without imposing any restriction on the initial data given
 on one half of the hourglass with the exception, of course, of the Gauss law constraint. This is because every
  differentiation of \eqref{psi3} in time brings coefficients of the transverse fields of one additional order.
   That is, $\Psi_{(n)}^a$ includes $\pi^a_{(n)}$, $f_{ar}^{(n)}$ and terms of smaller order.
    The longitudinal coefficients $\pi^r_{(n)}$ and $f_{\vartheta\varphi}^{(n)}$, for $n\neq 0$ can be
     obtained from the Gauss law and the Bianchi identity. Therefore, the appriopriate initial data 
     are the transverse fields $\pi^r_{(0)}$ and $f_{ar}^{(0)}$, 
     together with the electric and magnetic BMS charges, on one half of an 
     hourglass.
\section{Dictionary for translation to usual light cone variables}

\label{sec:app3}

In this appendix, a dictionary between the asymptotic conditions in
null coordinates in the original work of BMS  \cit{Bondi:1962px,Sachs:1962zza},
and the asymptotic conditions in the hyperbolic foliation here introduced,
is established.

The asymptotic form of the metric in a null foliation takes the form
\[
ds^{2}=e^{2\beta}\frac{V}{r}du^{2}-2e^{2\beta}dudr+G_{ab}\left(dx^{a}-U^{a}du\right)\left(dx^{b}-U^{b}du\right)\,,
\]
where 
\begin{align*}
\beta & =-\frac{c\bar{c}}{4r^{2}}+O\left(r^{-4}\right)\,,\\
\frac{V}{r} & =-1+\frac{2m_{B}}{r}+O\left(r^{-2}\right)\,,\\
U^{z} & =\frac{\left(1+z\bar{z}\right)\left[\left(1+z\bar{z}\right)\partial_{z}\bar{c}-2\bar{z}\mathit{\bar{c}}\right]}{2r^{2}}+\frac{2}{3r^{3}}\left\{ \left(1+z\bar{z}\right)\bar{c}\left[\left(1+z\bar{z}\right)\partial_{\bar{z}}c-2zc\right]-N^{z}\right\} +O\left(r^{-4}\right)\,,\\
G_{zz} & =\frac{4c}{\left(1+z\bar{z}\right)^{2}}r+\frac{g_{zz}^{\left(-1\right)}}{r}+O\left(r^{-2}\right)\,,\\
G_{z\bar{z}} & =\frac{2r^{2}}{\left(1+z\bar{z}\right)^{2}}+\frac{4c\bar{c}}{\left(1+z\bar{z}\right)^{2}}+\frac{g_{z\bar{z}}^{\left(-1\right)}}{r}+O\left(r^{-2}\right)\,.
\end{align*}

Here $m_{B}$, $c$, $\bar{c}$, $N^{z}$ and $N^{\bar{z}}$ are functions
of the retarded time $u$ and the stereographic coordinates on the
sphere $z,\bar{z}$, with $z=e^{i}\cot\left(\frac{\theta}{2}\right)$.

One passes to the hyperbolic coordinate by performing the changes of coordinates
$u=x^{0}-r$, and also
\eqref{eq:coordtransf}. 
Denoting with dots over the coefficients $c$, $\bar c$  partial derivatives with respect to their argument $t_{\mbox{\tiny ret}}$, one obtains, 
\begin{align*}
g_{\rho\rho}= & \frac{\tau_{0}^{2}}{\rho^{2}}-\left(\tau_{0}^{2}+\frac{1}{2}c\bar{c}\right)\frac{1}{\rho^{4}}-\tau_{0}\left[\left(c\bar{c}\right)^{\cdot}-2m_{B}\right]\frac{1}{\rho^{5}}+O\left(\rho^{-6}\right)\,,\\
g_{\rho z}= & -\frac{\tau_{0}\left[\left(1+z\bar{z}\right)\partial_{\bar{z}}c-2zc\right]}{2\rho^{2}\left(1+z\bar{z}\right)}\\
 & -\frac{8N^{\bar{z}}+\left(1+z\bar{z}\right)\left[3\tau_{0}^{2}\left(\left(1+z\bar{z}\right)\partial_{\bar{z}}\dot{c}-2z\dot{c}\right)+4c\left(\left(1+z\bar{z}\right)\partial_{z}\bar{c}-2\bar{z}\bar{c}\right)\right]}{12\left(1+z\bar{z}\right)^{2}\rho^{3}}+O\left(\rho^{-4}\right)\,,\\
g_{zz}= & \frac{4\tau_{0}c}{\left(1+z\bar{z}\right)^{2}}\rho+\frac{2\tau_{0}^{2}\dot{c}}{\left(1+z\bar{z}\right)^{2}}+\left(\frac{\tau_{0}^{3}\ddot{c}}{2\left(1+z\bar{z}\right)^{2}}+\frac{g_{zz}^{\left(-1\right)}}{\tau_{0}}\right)\frac{1}{\rho}+O\left(\rho^{-2}\right)\,,\\
g_{z\bar{z}}= & \frac{2\tau_{0}^{2}\rho^{2}}{\left(1+z\bar{z}\right)^{2}}+\left[\frac{2\tau_{0}\left(c\bar{c}\right)^{.}}{\left(1+z\bar{z}\right)^{2}}+\frac{g_{z\bar{z}}^{\left(-1\right)}}{\tau_{0}}\right]\frac{1}{\rho}+O\left(\rho^{-2}\right)\,,
\end{align*}

\begin{align*}
\pi^{\rho\rho}= & -\sqrt{\gamma}\rho^{3}-\frac{1}{2}\sqrt{\gamma}\left(1+\frac{c\bar{c}}{\tau_{0}^{2}}\right)\rho\\
 & +\frac{\sqrt{\gamma}}{8\tau_{0}^{3}}\left\{ \tau_{0}^{2}\left[\left(1+z\bar{z}\right)\left(-2z\partial_{\bar{z}}c+\left(1+z\bar{z}\right)\partial_{\bar{z}}^{2}c-2\bar{z}\partial_{z}\bar{c}+z\bar{z}\partial_{z}^{2}\bar{c}+\partial_{z}^{2}\bar{c}\right)+2\bar{c}\left(\bar{z}^{2}-\dot{c}\right)-4m_{B}\right]\right.\\
 & \left.+2\tau_{0}^{2}c\left(z^{2}-\dot{\bar{c}}\right)+2\left(1+z\bar{z}\right)^{2}g_{z\bar{z}}^{\left(-1\right)}\right\} +O\left(\rho^{-1}\right)\,,\\
\pi^{\rho z}= & -\frac{\sqrt{\gamma}}{6\tau_{0}^{2}}\left[\left(1+z\bar{z}\right)\bar{c}\left(\left(1+z\bar{z}\right)\partial_{\bar{z}}c-2zc\right)-N^{z}\right]\frac{1}{\rho^{2}}+O\left(\rho^{-3}\right)\,,\\
\pi^{zz}= & \frac{3\sqrt{\gamma}\left(1+z\bar{z}\right)^{2}\bar{c}}{4\tau_{0}\rho^{2}}+\frac{\sqrt{\gamma}\left(1+z\bar{z}\right)^{2}\dot{\bar{c}}}{2\rho^{3}}+O\left(\rho^{-4}\right)\,,\\
\pi^{z\bar{z}}= & -\frac{\sqrt{\gamma}\left(1+z\bar{z}\right)^{2}}{2\rho}+\frac{\sqrt{\gamma}\left(1+z\bar{z}\right)^{2}\left(2\tau_{0}^{2}-5c\bar{c}\right)}{8\tau_{0}^{2}\rho^{3}}\\
 & +\frac{\sqrt{\gamma}\left(1+z\bar{z}\right)^{2}}{32\tau_{0}^{3}\rho^{4}}\left\{ \tau_{0}^{2}\left[\left(1+z\bar{z}\right)\left(\partial_{z}^{2}\bar{c}\left(1+z\bar{z}\right)-2\bar{z}\partial_{z}\bar{c}+\left(1+z\bar{z}\right)\partial_{\bar{z}}^{2}c-2z\partial_{\bar{z}}c\right)\right.\right.\\
 & +\left.\left.\left(2\bar{z}^{2}-13\dot{c}\right)\bar{c}+\left(2z^{2}-13\dot{\bar{c}}\right)c+2m_{B}\right]+6\left(1+z\bar{z}\right)^{2}g_{z\bar{z}}^{\left(-1\right)}\right\} +O\left(\rho^{-4}\right)\,.
\end{align*}

As particular cases we have, 
\[
f_{zz}^{\left(1\right)}=\frac{4c}{\tau_{0}\left(1+z\bar{z}\right)^{2}}\quad,\quad f_{\bar{z}\bar{z}}^{\left(1\right)}=\frac{4\bar{c}}{\tau_{0}\left(1+z\bar{z}\right)^{2}}\,,
\]
\[
h_{\left(1\right)}^{zz}=\frac{1}{2}\sqrt{\gamma}\left(1+z\bar{z}\right)^{2}\dot{\bar{c}}\quad,\quad h_{\left(1\right)}^{\bar{z}\bar{z}}=\frac{1}{2}\sqrt{\gamma}\left(1+z\bar{z}\right)^{2}\dot{c}\,.
\]

The density associated to supertranslations ${\cal P}$ given in the
main text is related to $m_{B}$ according to
\[
{\cal P}=2\sqrt{\gamma}m_{B}.
\]
(The factor 2 appearing on the right-hand side of the above equation is a consequence of the choice
$8\pi G=1$ used here, in contradistinction with $16\pi G=1$ used
in  \cit{Bondi:1962px,Sachs:1962zza}.)

\end{document}